\documentclass[a4paper,fleqn,usenatbib]{mnras}

\usepackage{newtxtext,newtxmath}
\usepackage[T1]{fontenc}
\usepackage{ae,aecompl}

\usepackage{graphicx}	% Including figure files
\usepackage{amsmath}	% Advanced maths commands

\usepackage{amssymb}	% Extra maths symbols
\usepackage{subfigure}
\usepackage{enumitem} % for continued item numbering
\usepackage[normalem]{ulem}  % for strikethrough

\newcommand{\black}{\color{black}}

\newcommand{\ICM}{_{\mathrm{ICM}}}
\newcommand{\Part}{_{\mathrm{p}}}
\newcommand{\Max}{_{\mathrm{max}}}

\newcommand{\Kick}{_{\mathrm{kick}}}

\newcommand{\BV}{_{\mathrm{BV}}}
\newcommand{\Kepl}{_{\mathrm{K}}}

\usepackage[framemethod=tikz]{mdframed}

\title[Sloshing toy model]{A toy model for gas sloshing in galaxy clusters}

\author[E. Roediger et al.]{
E. Roediger,$^{1}$\thanks{E-mail: e.roediger@hull.ac.uk (ER)}
I. Vaezzadeh$^{2,1}$,
and P. Nulsen$^{3,4}$
\\
$^{1}$E.A. Milne Centre, University of Hull, Cottingham Road, HU6 7RX, U.K.\\
$^{2}$I. Physikalisches Institut, Universität zu Köln, Zülpicher Str. 77, D-50937 Köln, Germany\\
$^{3}$Center for Astrophysics|Harvard \& Smithsonian, 60 Garden St., Cambridge, MA 02138, USA\\
$^{4}$International Centre for Radio Astronomy Research, University of Western Australia, 35 Stirling Hwy, Crawley, WA 6009, Australia\\
}

\date{Accepted XXX. Received YYY; in original form ZZZ}

\pubyear{2023}

\begin{document}
\label{firstpage}
\pagerange{\pageref{firstpage}--\pageref{lastpage}}
\maketitle

\begin{abstract}
We apply a toy model based on `pendulum waves' to gas sloshing in galaxy clusters.
Starting with a galaxy cluster potential filled with a hydrostatic intra-cluster medium (ICM), we perturb all ICM by an initial small, unidirectional velocity, i.e.,~an instantaneous kick.
Consequently, each parcel of ICM will oscillate due to buoyancy with its local Brunt-Väisälä (BV) period, which we show to be approximately proportional to the cluster radius.
The oscillation of gas parcels at different radii with different periods leads to a characteristic, outwards-moving coherent pattern of local compressions and rarefactions; the former form the sloshing cold fronts (SCFs).
Our model predicts that SCFs (i) appear in the cluster centre first, (ii) move outwards on several Gyr timescales, (iii) form a staggered pattern on opposite sides of a given cluster, (iv) each move outwards with approximately constant speed; and that (v) inner SCFs form discontinuities more easily than outer ones.
These features are well known from idealised (magneto)-hydrodynamic simulations of cluster sloshing.
We perform comparison hydrodynamic+N-body simulations where sloshing is triggered either by an instantaneous kick or a minor merger.
Sloshing in these simulations qualitatively behaves as predicted by the toy model.
However, the toy model somewhat over-predicts the speed of sloshing fronts, and does not predict that inner SCFs emerge with a delay compared to outer ones.
In light of this, we identify the outermost cold front, which may be a `failed' SCF, as the best tracer of the age of the merger that set a cluster sloshing.
%It should be a single paragraph not more than 250 words (200 words for Letters).
%No references should appear in the abstract.
\end{abstract}
\begin{keywords}
Galaxies: clusters: intracluster medium ---
Galaxies: clusters: general ---
Physical Data and Processes: hydrodynamics
\end{keywords}

%%%%%%%%%%%%%%%%%%%%%%%%%%%%%%%%%%%%%%%%%%%%%%%%%%%%%%%%%%%%%%%%%%%%%%%%%%%%%%%%%%%%%%%%%%%%%%%%%%%%
%%%%%%%%%%%%%%%%%%%%%%%%%%%%%%%%%%%%%%%%%%%%%%%%%%%%%%%%%%%%%%%%%%%%%%%%%%%%%%%%%%%%%%%%%%%%%%%%%%%%
%%%%%%%%%%%%%%%%% BODY OF PAPER %%%%%%%%%%%%%%%%%%

\black
%***********************
\section{Introduction}
%***********************
Mergers between galaxy clusters leave observable features in the X-ray emitting intracluster medium (ICM) of a galaxy cluster.
These features include shocks and cold fronts \citep{Markevitch2007}.
Cold fronts differ from shocks in that the pressure is continuous across a cold front, so that the denser side of the discontinuity is colder than the more diffuse side.

Here we focus on sloshing cold fronts (SCFs), which arise when the ICM of a cluster is perturbed by, e.g., a minor merger as first proposed by \citet{Tittley2005,Ascasibar2006}.
They showed that the gravitational disturbance caused by a subcluster passing through the primary cluster is sufficient to cause the ICM of the primary cluster to `slosh' about the gravitational potential minimum, leading to the familiar arc-shaped `edges' in X-ray surface brightness, wrapped around the cluster core.
SCFs have been observed in many galaxy clusters (for a review see \citealt{ZuHone2016}) and are thought to be ubiquitous in cool-core (CC) clusters \citep{Markevitch2003,Ghizzardi2010}.

The development and evolution of SCFs has been well-studied using hydrodynamic simulations both in the interest of constructing cluster merger histories \citep{Roediger2011,Roediger2012fastslosh,Su2017sloshing,Sheardown2018,Vaezzadeh2022} and constraining transport processes within the ICM \citep{ZuHone2011,Roediger2013virgovisc,ZuHone2013,ZuHone2015,Brzycki2019a}.
\citet{Keshet2023} describe a spiral structure as a quasi-stationary solution for the ICM.

The positive entropy gradient in the ICM leads to the ICM being stable against convection, i.e.~after a perturbation a parcel of ICM oscillates rather than keeping rising or sinking.
The frequency of such a radial oscillation is known as the Brunt-Väisälä frequency \citep{Cox1980}, and can be written as:
\begin{equation}
\label{eqn:BVeqn}
    \omega_{\mathrm{BV}}(r)=\Omega_{\mathrm{K}} \sqrt{\frac{1}{\gamma}  \frac{d \ln K(r)}{d \ln r}}
\end{equation}
where $\Omega_{\mathrm{K}}=\sqrt{\frac{GM}{r^3}}$ is the Keplerian frequency, $\gamma=5/3$ is the ratio of specific heats and $K=kTn^{-2/3}$ the entropy index.
\citet{Churazov2003} and \citet{Su2017sloshing} have used the BV period,
\begin{equation}
\label{eq_BVperiod}
    T\BV=2\pi/\omega\BV 
\end{equation}
as an estimate of the sloshing timescale.

In this paper we present a toy model that links local oscillations of ICM parcels with their BV period to the global motion of SCFs, following in broad terms the scenario outlined in \citet{Churazov2003}.
In essence, our toy model draws an analogy between sloshing fronts and pendulum waves (e.g.~\citealt{Flaten2001}).
To this end, in Section~\ref{sec_toy} we describe the basic toy model, summarising its predictions in Section~\ref{sec_toyresults1}.
Section~\ref{sec_toyAdvanced} discusses extensions to the basic toy model, in particular variations of the initial perturbation.
Section~\ref{sec_hydro} introduces the hydrodynamic simulations of sloshing resulting from an instantaneous kick and from a minor binary cluster merger, and compares the motion of the sloshing fronts in the simulations with the toy model predictions.
Sections~\ref{sec_discussion} and \ref{sec_summary} discuss limitations and implications, and summarise the results, respectively.
A more detailed account of the phenomenology of sloshing in terms of the linear perturbations of a cluster atmosphere is presented in Nulsen et al.~(in prep.).

%*********************************************************
\section{The basic toy model}
\label{sec_toy}
\begin{figure}
    \includegraphics[width=\columnwidth]{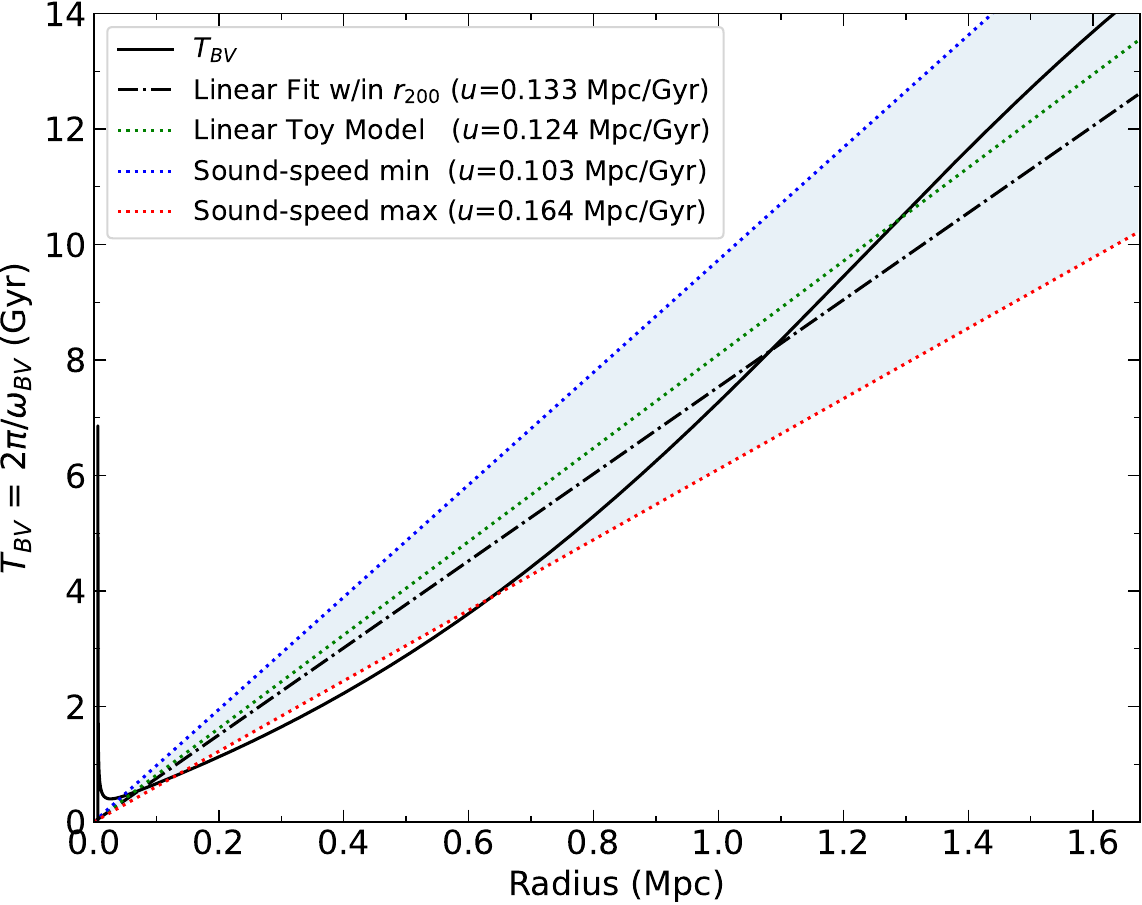}
    \caption{Agreement of the correct sloshing timescale for our model cluster used in Section~\ref{sec_hydro} with the simple approximation from Equation~\ref{eq_T_BV_linear}.
    	Solid black line: Sloshing timescale calculated from the BV period (Equation~\ref{eqn:BVeqn}) for our M$_{200}=5\times10^{14}M_{\odot}$ cluster as a function of its radius out to $r_{200}$ (1.67\,Mpc).
	Dash-dotted black line: A linear regression fitted to the sloshing timescale which yields a characteristic sloshing front speed of $u=0.133$\,Mpc/Gyr (see Equations~\ref{eq_oscillatorSpeeds} and \ref{eq:TofR}).
	Dotted green line: the linear approximation from Equation~\ref{eq_T_BV_linear} assuming a single sound speed within $r_{200}$, which yields a sloshing speed of $u=0.124$\,Mpc/Gyr.
	Dotted blue and red lines and shaded region: the linear approximation from Equation~\ref{eq_T_BV_linear} using the maximum and minimum values of the sound speed within $r_{200}$.}
\label{fig:P_slosh}
\end{figure}

%*********************************************************
As a first step whose result is needed later, we show that the BV period, $T\BV$, in a galaxy cluster is approximately a linear function of radius.
To this end, we write the BV period (Equation~\ref{eq_BVperiod}) using the Kepler speed $v\Kepl = \Omega\Kepl r$:
%=======
\begin{equation}
    T\BV(r)= 2\pi \; v\Kepl^{-1} \gamma^{1/2} \left(\frac{d \ln K(r)}{d \ln r}\right)^{-1/2} r.
\end{equation}
%=========
In galaxy clusters, it is established empirically that both the Kepler speed $v_\mathrm{K}$ and the logarithmic derivative of the entropy index $\frac{d \ln K(r)}{d \ln r}$ are approximately constant with radius.
If $K$ is a power law $K\propto r^q$, its logarithmic derivative is its power index $q$.
Theoretically derived and observed entropy power law indices are $1.1$ to $1.2$ \citep{Tozzi2001,Voit2005,Cavagnolo2009}.
Thus, we can write the BV period as
%=======
\begin{eqnarray}
T_{\mathrm{BV}}(r) &=& \frac{1}{u}r \;\;\textrm{with} \label{eq:TofR} \label{eq_u_final} \label{eq_T_BV_linear}\\
 u &=& \frac{1}{2\pi} \sqrt{ \frac{q}{\gamma} } v_\mathrm{K} \approx 0.13v_\mathrm{K}\;\;\textrm{for}\;\;q=1.1 \nonumber\\
 \textrm{or}\;\;\; u &\approx& 0.15c_s. \nonumber
\end{eqnarray}
%=========
In the last step we made use of the fact that in a hydrostatic cluster, the Kepler speed $v\Kepl$ is comparable to the sound speed $c_s$.
We write the proportionality constant as $1/u$, as the quantity $u$ will turn out to be the characteristic sloshing front speed.
Later in this paper we present hydrodynamic sloshing simulations for a model cluster.
In Figure~\ref{fig:P_slosh}, we compare the BV period of our model cluster, calculated by the full equation (Equation~\ref{eqn:BVeqn}), with the approximation from Equation~\ref{eq_u_final}.

%FFFFFFFF
%FFFFFFF

%*************
\subsection{A row of simple harmonic oscillators whose period depends on the position of their equilibrium points}
Our model starts with a galaxy cluster potential filled with a hydrostatic ICM.
In the simplest version, we imagine the ICM to be perturbed by an instantaneous, unidirectional `kick', i.e.~all ICM is given an initial small velocity in the same direction.
For the sake of the toy model, we first focus on the resulting motion of ICM parcels along the cluster radius, $r$, parallel to the kick direction, such that the kick is directed outwards, in positive $r$-direction.
As a result of the kick, each ICM parcel along this radius will oscillate radially due to buoyancy with its local Brunt-Väisälä (BV) period (Equation~\ref{eq_T_BV_linear}).
Due to the radial variation of the oscillation period, patterns of compression and rarefaction regions will arise.
In this toy model, we assume that the ICM parcels simply oscillate locally without influencing each other like in a pendulum wave experiment (\citealt{Flaten2001}), which is a simplification of the fluid nature of the ICM.
 
To visualise the resulting patterns, we imagine a row of simple harmonic oscillators along this radius.
To start with, we assume that each oscillator is kicked such that all oscillators have the same positive amplitude,~$A$.
Variations to this perturbation are discussed below.

The displacement of each oscillator away from its equilibrium position at time, $t$, shall be $D(t)$; its period is the BV period, assumed to depend linearly on radius as shown in Equation~\ref{eq_T_BV_linear}.
Thus, we can write the displacement, $D$, away from equilibrium of the oscillator with equilibrium position, $r$, at time, $t$, as
%==========
\begin{equation}
\label{eq:oscillatorDisplacement}
    D(r,t) = A \sin\left( \frac{2\pi}{T\BV(r)} t +\phi  \right).
\end{equation}
%==========
If indeed all oscillators receive their first kick at the same time, $t=0$, the phase is zero ($\phi=0$).
In a more general case, the different oscillators could start at different moments in time, and thus have different phases; we discuss this case below.

\begin{figure}
    \includegraphics[width=\columnwidth]{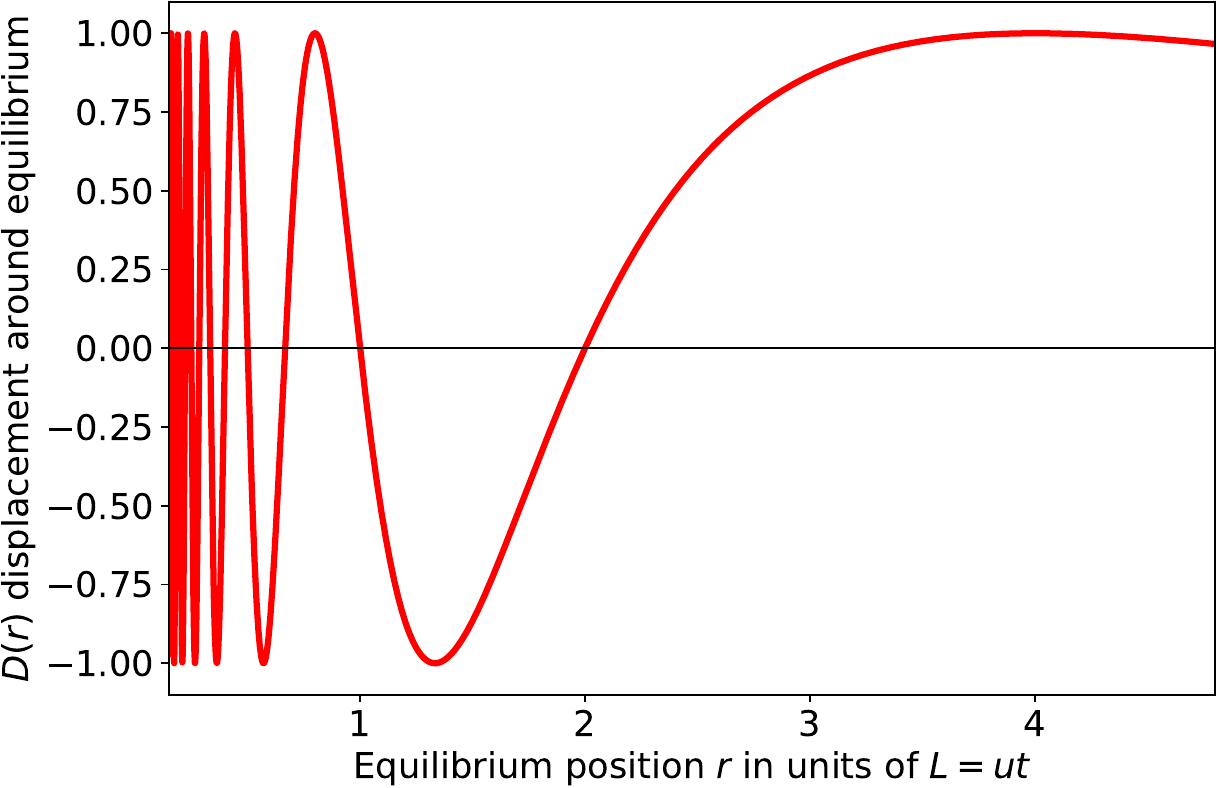}	
    \caption{Displacement, $D$, away from the equilibrium position of oscillators as a function of their equilibrium position, $r$ (see Equation~\ref{eq:oscillatorDisplacement2}).}
\label{fig:displacement}
\end{figure}

With Equation~\ref{eq:TofR}, the displacement, $D$, of the oscillator with equilibrium position, $r$, at time, $t$, becomes
%==========
\begin{equation}
\label{eq:oscillatorDisplacement2}
    D(r,t) = A \sin\left( \frac{2\pi u}{r} t  \right) = A \sin\left( \frac{2\pi L}{r}  \right) \;\; \mathrm{with} \;L=ut;
\end{equation}
%==========
its dependence on radius for a fixed time, $t$, is shown in Figure~\ref{fig:displacement}.

%FFFFFFFFFFFFF
%FFFFFFFFFFFFF

The $1/r$ term in the argument of the sine function distorts the sine function such that its `wavelength' decreases with decreasing $r$.
The displacement $D(r)$ approaches zero for $r>4L$, and its outermost $x$-axis intercept is at $r=2L$.

%*************
\subsection{Relating sloshing cold fronts to the row of oscillators} \label{sec_CF_oscillators}
Once the oscillators start their oscillation, the variation in period along $r$ will lead to a pattern of enhanced, and reduced, densities of the oscillating parcels.
 The locations of enhanced densities of oscillating parcels mark the locations of actual SCFs.
These locations occur close to those $r$-intercepts (zeros) of $D(r)$ where $D(r)$ has a negative slope.
The zeros can be found easily by setting the argument of the sine function to multiples of $\pi$, i.e.~the zeros occur at
%==========
\begin{equation}
r_n = \frac{2L}{n}, \;\;n=1,2,3,\ldots, \nonumber
\end{equation}
%==========
but only the even-numbered ones are those with a negative slope.
The oscillators with equilibrium points just left of these zeros have moved towards the right (outwards), and the oscillators with equilibrium points just right of these zeros have moved towards the left (inwards).
Consequently, the density of oscillators is enhanced around those zeros.
This pattern of alternating inwards and outwards motion is well-known in hydrodynamic simulations of sloshing; the SCFs are located where outward-moving ICM meets inward-moving ICM, as described above.

\begin{figure*}
	\includegraphics[width=\columnwidth]{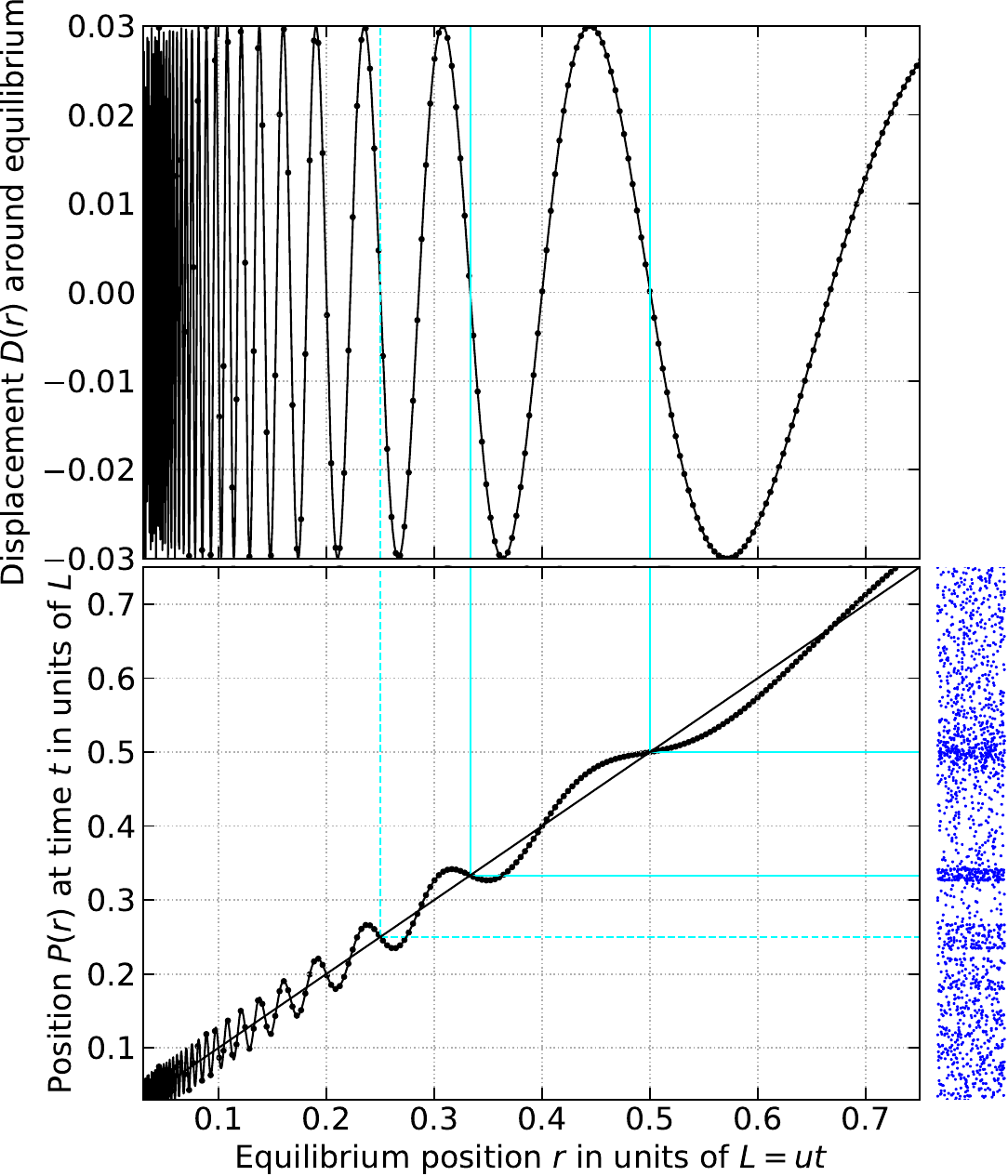}	 
	\includegraphics[width=\columnwidth]{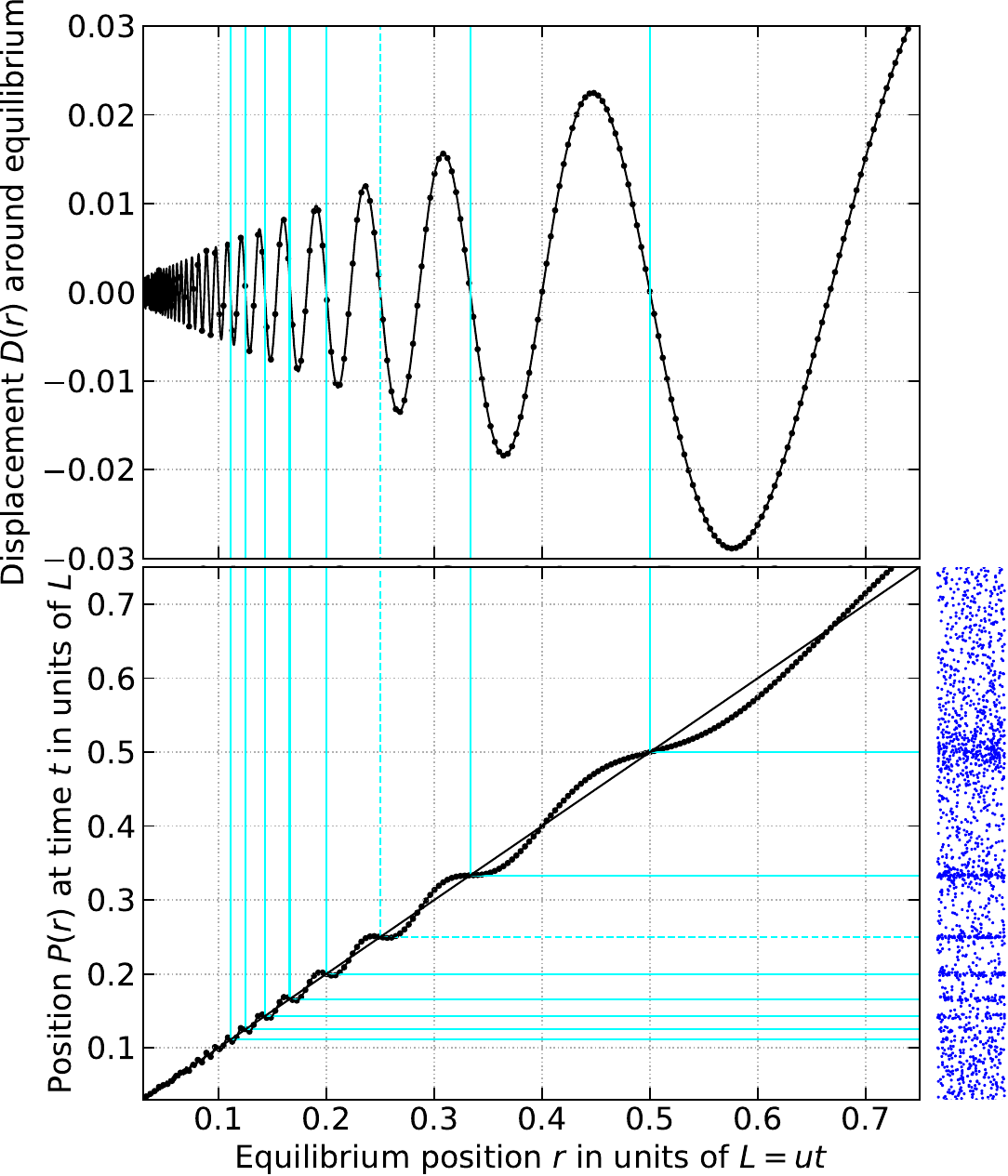}	
	\caption{Oscillator positions and enhanced densities.
		The left-hand-side panels are for the case of constant oscillator amplitude, the right-hand-side panels illustrate the case where the amplitude grows linearly with radius.
		The top panels show the displacement $D(r)$ of the oscillators around their equilibrium position (compare to Figure~\ref{fig:displacement}).
		The bottom panels show the distance of the oscillator positions to the cluster centre at a given time.
		The bar of blue dots on the right in the bottom panels visualises the enhanced densities of oscillators at certain radii, see text for full description.
		The cyan lines in all panels mark locations of potential sloshing cold fronts, i.e.~locations of enhanced densities of oscillators, i.e.~locations of zeros of $D(r)$ with negative slopes.}
    \label{fig:positions}
\end{figure*}

A second method to visualise the enhanced density of oscillators is to consider the actual positions of the oscillators at a given time~$t$.
The position of the oscillator with equilibrium position, $r$, at time, $t$, with respect to the cluster centre is
%==========
\begin{equation}\
\label{eq:oscillatorPosition}
    P(r,t) = r+D(r,t) = r + A \sin\left( \frac{2\pi u}{r} t \right),
\end{equation}
%==========
i.e.~its equilibrium position plus its local displacement.
We show the function $P(x)$ along with $D(x)$, both zoomed into a relevant range of $r$, in Figure~\ref{fig:positions} in the left-hand-side column.

%FFFFFFFFFFFFF
%FFFFFFFFFFFFF

SCFs will occur where the density of oscillating parcels is enhanced.
 This is at the radii where the slope of $P(r)$ has local minima, which is at, or near, the zeros of $D(r)$ with negative slopes.
These radii are marked by cyan lines in Figure~\ref{fig:positions}.
However, at smaller radii, neighboring enhanced regions start overlapping, which washes out the enhancements again.
This would be apparent if we took the black markers in the bottom left panel of Figure~\ref{fig:positions} (which are equidistant in their equilibrium position) and project them onto the $P$-axis.
The result of such a projection is shown by the band of blue dots on the right of the panel.
Instead of projecting the black markers directly onto the $P$-axis, i.e.~setting all their $r$-coordinates to zero, we projected them into a small $r$-range in the margin of the plot, giving each a random $r$-coordinate in that range to avoid crowding.
There are bands of clearly enhanced densities: these mark the locations of SCFs.

As stated above, due to overlap at inner radii, there are only a limited number of regions with actual enhanced oscillator density.
This washing-out of inner SCFs will be reduced with a more realistic initial perturbation, as discussed below.
As real ICM parcels cannot cross through each other, this washing-out effect might not happen in reality.
The outermost SCFs do not suffer from crowding and should always exist, though they may not be discontinuities, as discussed below.

%*************
\subsection{A more realistic perturbation: constant initial velocity instead of amplitude}
The velocity amplitude of the oscillation stated in Equation~\ref{eq:oscillatorDisplacement2} would be
\[v_{\max} = \frac{2\pi A u}{r},\]
i.e.~it would depend on $r$ because the oscillation period depends on $r$.
This would mean oscillators at lower $r$ would have a much higher energy if they had equal masses.
A more even energy distribution with radius would be more realistic for the ICM of a cluster.
Under the impulse approximation, i.e.~assuming the perturber passes a region faster than the matter can respond, the perturbation caused is a constant velocity kick if the perturber is an isothermal sphere.

To achieve a constant velocity amplitude, the oscillation amplitude, $A$, needs to be a matching function of radius, $r$:
%i.e., equal specific energy for all oscillators,  
%==========
\begin{equation}
    A=A(r)=\alpha r,
\end{equation}
%==========
where $\alpha$ is a positive dimensionless parameter (signifying an outwards kick), and consequently the velocity amplitude is constant:
%==========
\begin{equation}
\label{eq_vkick}
    v_{\max} = 2\pi \alpha u \;\;\;\textrm{or}\;\;\; \alpha=\frac{v\Kick}{2\pi u}
\end{equation}
%==========
The above relationship links the proportionality constant $\alpha$ for $A(r)$ to the initial kick velocity if we want to think of a scenario where the initial condition is a constant speed for all oscillators instead of a constant amplitude.
Thus, the position of each oscillator now is
%==========
\begin{equation}
\label{eq:oscillatorPosition_vkickConst}
    P(r,t) = r + \alpha r \sin\left( \frac{2\pi u}{r} t  \right) = r + \alpha r \sin\left( \frac{2\pi L}{r} \right) \;\;\textrm{with}\;\;L=ut.
\end{equation}
%==========
This new displacement, $D(r)$, and new oscillator position, $P(r)$, are shown in the right-hand-side column of Figure~\ref{fig:positions}.
Again, the oscillators will pile up where the slope of $P(r)$ has its local minima.
Differentiating $P(r)$ with respect to $r$ yields its slope,
%==========
\begin{equation}
    \frac{\partial P}{\partial r} = 1 + \alpha \sin\left( \frac{2\pi L}{r} \right) - \frac{ 2\pi\alpha L}{r} \cos\left( \frac{2\pi L}{r} \right), \nonumber
\end{equation}
%==========
differentiating again to find the minima of the slope yields
%==========
\begin{equation}
    \frac{\partial^2 P}{\partial r^2} = - \frac{4\pi^2L^2\alpha}{r^3} \sin\left( \frac{2\pi L}{r} \right), \nonumber
\end{equation}
%==========
which has zeros where the argument of the sine function equals integer multiples of $\pi$.
Thus, minima and maxima of the slope of $P(r)$ occur again at 
%==========
\begin{equation}
    r_n = \frac{2L}{n}, \;\;n=1,2,3,\ldots, \nonumber
\end{equation}
%==========
but only the even-numbered instances are minima, i.e.~locations of sloshing fronts.
We mark them with cyan lines in Figure~\ref{fig:positions}.

In the blue-dot-band next to the bottom-right panel of Figure~\ref{fig:positions}, we repeat the exercise of visualising the enhanced oscillator densities at SCFs.
Making the amplitude a linear (or, more generally, monotonically growing) function of $r$ strongly reduces the washing out of inner SCFs.
In the shown example, 7 to 8 SCFs can be identified instead of only 3 fronts in the left-hand-side panels.

%*************
\subsection{The location and motion of sloshing cold fronts}
We established that along the radius along which the initial kick was directed outwards, SCFs appear at the even-numbered instances of
%==========
\begin{equation}
\label{eq_oscillatorRoots}
    r_{n} = \frac{2L}{n} = \frac{2u}{n}t ;\; \mathrm{for} \;n=1,2,3,\ldots.
\end{equation}
%==========
To consider the case of a kick in the negative $r$-direction, i.e.~the other side of the cluster in the case of a cluster-wide unidirectional kick, we need to invert the sign of the amplitude, $A$, in Equations~\ref{eq:oscillatorDisplacement2} and \ref{eq:oscillatorPosition}, and of the parameter $\alpha$ in Equation~\ref{eq:oscillatorPosition_vkickConst}.
Finding again the locations of minimum slope in $P(r)$ now identifies the odd-numbered instances of Equation~\ref{eq_oscillatorRoots}.

Thus, Equation~\ref{eq_oscillatorRoots} lists the locations of potential SCFs on both sides of the cluster, counting SCFs from the outermost one inwards, alternating sides of the cluster.
The sloshing fronts form a staggered pattern.
The outermost front is expected on the side of the cluster that experienced the inwards kick.

Each SCF moves outwards with a constant speed,
%==========
\begin{equation}
\label{eq_oscillatorSpeeds}
    v_{n} = \frac{2u}{n} ;\; \mathrm{for} \;n=1,2,3,\ldots,
\end{equation}
%==========
the inner fronts move slower than outer ones, and the pattern of $D(r)$ remains self-similar.
As the characteristic speed, $u$, is much smaller than the sound speed, all SCFs move subsonically.
We note that in the toy model framework, sloshing cold fronts are simply a pattern of local enhancements that is travelling through space, similarly to the patterns seen in a pendulum wave experiment (\citealt{Flaten2001}).

%*****
\subsection{True and `failed' sloshing fronts}
At the locations of the potential cold fronts identified by Equation~\ref{eq_oscillatorRoots}, the slope of the function $P(r)$ decreases with decreasing $r$.
For the outermost potential fronts, the slope of $P(r)$ can still be positive, but for more inner potential fronts it is negative.
In the framework of the toy model, a region of negative $\partial P/\partial r$ means oscillators have crossed through each other, whereas at the outermost fronts where the slope of $P(r)$ is positive, the oscillators have not changed their order but simply moved closer together.
A profile of oscillator density as a function of radius shows a continuous enhancement at an outer potential front when $\partial P/\partial r>0$, but the oscillator density shows a discontinuous enhancement if $\partial P/\partial r<0$ (see Figure~\ref{fig:positions}).
It is known from hydrodynamical sloshing simulations that for mild mergers the outermost sloshing `fronts' can fail to become discontinuities, whereas the inner fronts are discontinuous.
Thus, by analogy, in the toy model we identify potential sloshing fronts with $\partial P/\partial r>0$ as `failed' fronts.
True, discontinuous fronts require $\partial P/\partial r<0$.

We note that in a 1-D scenario, ICM parcels would not pass through each other, and with adiabatic processes alone the gaseous ICM would not form discontinuities.
However, we know from observations and hydrodynamic simulations that in 3-D discontinuous fronts form.
The toy model alone cannot explain the exact process, though.
Within the toy model framework, the formation of true, discontinuous fronts occurs where the slope of $P(r)$ not only has a local minimum, but the slope is zero or negative at that minimum.
Equivalent considerations for both sides of the cluster lead to the following condition for true, discontinuous fronts:
%==========
\begin{eqnarray}
\label{eq_vkick_strength}
    \frac{\partial P}{\partial r} (r_n) = 1-\pi\alpha n = 1-\frac{v\Kick}{u} \frac{n}{2} &\le& 0 \;\; \textrm{or} \nonumber \\
    \frac{2u}{v\Kick} &\le& n.
\end{eqnarray}
%==========
This is progressively easier to fulfil for potential fronts with higher $n$, i.e.~closer to the cluster centre.
If this condition is not fulfilled, we expect the SCF to not be a discontinuity but only a gradient in ICM density and temperature in the correct direction, and we call this a `failed' SCF.

%*************
\subsection{Predicted behaviour of sloshing cold fronts from basic toy model} \label{sec_toyresults1}
In summary, our toy model made the following assumptions:
\begin{itemize}
    \item We considered the ICM along one diameter in an initially hydrostatic cluster.
    \item All ICM parcels along this diameter simultaneously receive a kick, i.e.~a small, unidirectional, initial velocity.
    \item As a result, the ICM parcels will oscillate locally as simple harmonic oscillators along the diameter around their equilibrium radius with their local BV period. Their amplitude shall be small compared to the radial range of interest (this defines `small kick' in the assumption above).
    \item The BV period depends linearly on cluster radius, see Equation~\ref{eq_T_BV_linear}.
\end{itemize}

We showed that the dependence of the oscillation period on radius leads to density enhancements appearing in the ICM.
We identified those as (potential) SCFs.

Based on these assumptions, this toy model {predicts the following behaviour of SCFs}:
\begin{enumerate}
    \item The radii of SCFs on opposite sides of the cluster make a staggered pattern.
    \item At any given time, SCFs are located at the radii given in Equation~\ref{eq_oscillatorRoots} (the index counts inwards), i.e.~their radii keep a self-similar pattern over time.
    \item Each SCF moves outwards with constant, clearly subsonic speed (Equations~\ref{eq_oscillatorSpeeds} and \ref{eq_u_final}).
    \item Inner SCFs move slower than outer ones (Equation~\ref{eq_oscillatorSpeeds}).
    \item For every time, $t$, there is an outermost SCF, i.e.~despite the assumed cluster-wide perturbation, the SCF pattern will grow from the cluster centre outwards.
    \item Not all sloshing fronts identified by Equation~\ref{eq_oscillatorRoots} are true fronts.
    	In particular the outer `fronts' could fail to become true discontinuities.
	Forming true discontinuities requires a sufficiently strong initial kick velocity as specified in Equation~\ref{eq_vkick_strength}.
	This condition is easier to fulfil for inner sloshing fronts.
	However, even a very mild kick velocity will lead to a sloshing-front-like pattern, except that the classic discontinuities are replaced by corresponding slopes in density and temperature.
\end{enumerate}
The qualitative aspects of these predictions are well known SCF features in hydrodynamic simulations.

%*********************************************************
\subsection{Variations to the toy model}
\label{sec_toyAdvanced}
%*********************************************************
An obvious question is whether the chosen initial perturbation impacts the prediction.
To this end, we discuss some variations to the initial perturbation, namely:
\begin{itemize}
    \item an initial offset instead of an initial velocity,
    \item a constant oscillation amplitude throughout the cluster instead of a constant oscillator velocity or energy,
    \item a non-simultaneous perturbation where the perturber's velocity through the cluster is much faster than the characteristic speed $u$ identified above.
\end{itemize}
This section reveals that all qualitative conclusions are unaffected, and even quantitative results for sloshing front locations, and speeds, are very similar.
The strongest impact could arise from the second point.
The constant oscillator velocity throughout the cluster favours the appearance of numerous true fronts, whereas in the constant amplitude case inner cold fronts could be washed out by overlapping each other, although the toy model cannot predict how gas parcels would behave in this scenario.

%*************
\subsubsection{Initial offset instead of initial kick}
If we consider the case of an initial offset instead of an initial kick, the equation describing the position of each oscillator as a function of its equilibrium position (equivalent of Equation~\ref{eq:oscillatorPosition_vkickConst}) becomes
%==========
\begin{equation}
\label{eq:oscillatorPosition_Offset_vkickConst}
    P(r,t) = r + \alpha r \cos\left( \frac{2\pi u}{r} t \right).
\end{equation}
%==========
Here we have kept a radius-dependent initial offset or amplitude, i.e.~the maximum oscillation velocity (and energy) of each oscillator is the same.
Considerations equivalent to the ones above reveal that now potential sloshing cold fronts are expected at locations (equivalents to Equations~\ref{eq_oscillatorRoots} and \ref{eq_oscillatorSpeeds})
%==========
\begin{equation}
    r_{n} = \frac{2ut}{n+1/2} ;\; \mathrm{for} \;n=1,2,3,\ldots.
\end{equation}
%==========
and their speeds are
%==========
\begin{equation}
    v_{n} = \frac{2u}{n+1/2} ;\; \mathrm{for} \;n=1,2,3,\ldots.
\end{equation}
%==========
Again, sloshing fronts are numbered from the outermost one inwards.

Odd numbered fronts appear on the side where the offset perturbation was directed towards the cluster centre, even-numbered fronts on the other side.
The difference of front speeds between the two perturbation modes becomes less with increasing $n$, i.e.~for inner fronts.
For front number $n$ to be a true front, i.e.~an ICM discontinuity, the kick velocity needs to obey (equivalent of Equation~\ref{eq_vkick_strength})
%==========
\begin{equation}
    \frac{2u}{v\Kick} \le n+1/2.
\end{equation}
%==========
Thus, all qualitative conclusions remain, and quantitative conclusions change only mildly.

%*************
\subsubsection{Constant amplitude throughout cluster instead of constant velocity/energy}
If the initial perturbation would lead to oscillators at different radii having the same amplitude rather than the same energy, we would expect fewer true cold fronts.
Inner cold fronts would wash each other out easily (see the bottom left panel of Figure~\ref{fig:positions}).
In the case of constant oscillator energy throughout the cluster, the resulting radial growth of amplitude reduces fronts being washed out at smaller radii, and supports fronts being true fronts at larger radii.

%*************
\subsubsection{Non-simultaneous perturbation}
We return to the case of an initial kick.
So far we considered the case that the perturbation occurs at the same time throughout the cluster.
We relax this condition now by expressing the displacement of an oscillator with equilibrium position $r$ at time $t$ as 
%==========
\begin{equation}
\label{eq_oscillatorDisplacement_Delay}
    D(r,t) = A \sin\left( \frac{2\pi u}{r} [t-\tau(r)] \right),
\end{equation}
%==========
i.e.~we use a location-dependent delay time, $\tau(r)$.
For $t<\tau$, $D$ shall be zero.
As a simple case, we write the location-dependent delay time $\tau(r)$ as
%==========
\begin{equation}
    \tau(r) = \tau_0 - \frac{1}{v_p}r.
\end{equation}
%==========
This describes a perturber arriving at cluster radius $r\Max = v_p\tau_0$ at $t=0$, which takes the time, $\tau_0$, to travel to the cluster centre, travelling with constant speed $v_p$ inwards.

Inserting the delay function, $t(r)$, into the displacement function yields
%==========
\begin{equation}
\label{eq_oscillatorDisplacement_Delay2}
    D(r,t) = A \sin\left( 2\pi \left[ \frac{ u}{r} (t-\tau_0) + \frac{u}{v_p} \right] \right).
\end{equation}
%==========
The zeros, i.e.~the locations of potential sloshing fronts on alternating sides of the cluster, can be derived as above, and are
%==========
\begin{equation}
    r_{n} = \frac{2u (t-\tau_0) }{n-2u/v_p} ;\; \mathrm{for} \;n=1,2,3,\ldots,
\end{equation}
%==========
and their speeds are
%==========
\begin{equation}
    v_{n} = \frac{2u}{n-2u/v_p} ;\; \mathrm{for} \;n=1,2,3,\ldots.
\end{equation}
%==========
For a typical cluster, the velocity, $u$, characterising the dependence of BV period of radius, is only about 15\% of the sound speed (Equation~\ref{eq_u_final}), whereas the infall velocity of a subcluster, i.e.~a perturber, is easily 1.5 times the sound speed.
Thus, the characteristic sloshing front speed, $u$, is at least 10 times smaller than the typical speed of a perturber crossing the cluster, and $u/v_p$ is small.
Thus, if we shift into the time frame $\tilde t = t-\tau_0$ where the perturber arrives in the cluster centre at $\tilde t = 0$, the positions and speeds of the potential cold fronts are only slightly larger compared to the fully instantaneous perturbation approach.
The effect is largest (of the order of 10\%) for the outermost front.
All qualitative conclusions remain the same.
However, we note that this scenario still assumes a locally instantaneous perturbation, and not a perturbation over an extended amount of time or region.

%*************
\subsubsection{BV period not a linear function of radius}
The outward motion of SCFs will occur, even if not at constant speed, as long as the BV period is a monotonically increasing function of radius.
The positions and speeds of sloshing fronts can be calculated by the same formalism as above but may require a numerical solution.

%*********************************************************
\section{Comparison with hydrodynamical simulations}
\label{sec_hydro}
%*********************************************************
%*************
\subsection{Simulation Method}
In order to test the efficacy and predictive power of the toy model, we perform a set of three highly idealised simulations (dubbed Kick1, Kick2 and Kick3) in addition to an idealised binary merger simulation for comparison.
We initialise a spherically symmetric cluster (M$_{200} = 5 \times 10^{14} M_{\odot}$, r$_{200}$ = 1.67\,Mpc) in hydrostatic equilibrium.
Details of the method used to generate the cluster used in these simulations can be found in \citet{Vaezzadeh2022} which follows the methods of \citet{ZuHone2011}.
The particles are set up to form a dark matter (DM) halo, as explained in \citet{Vaezzadeh2022}, at rest in the grid, i.e the particles are not given any bulk velocity.
The gas in our simulation domain is initialised with a uniform initial velocity to the right which we vary between our three simulations.
This method is similar to the one used by \citet{Churazov2003} who used a planar shock front running over a cluster to initiated sloshing.
In simulations Kick1, Kick2, and Kick3, the gas has an initial velocity of 100\,km/s, 250\,km/s, and 500\,km/s respectively.
The cluster has a typical sound speed (calculated via $c_s = \sqrt{\gamma kT\ICM / m\Part}$, with $T\ICM = T_{200}$ = 2.78\,keV) of 863\,km/s (0.88\,Mpc/Gyr), which leads to a characteristic sloshing speed (via Equation~\ref{eq_u_final}) of 134\,km\,s$^{-1}$ (0.124\,Mpc/Gyr).

The simulations are run using the hydrodynamic + N-body code, FLASH v4.6 \citep{Fryxell2000}.
FLASH is an Eulerian adaptive mesh refinement (AMR) hydrodynamics code which allows us to save computational effort in areas of the simulation domain that are of little interest.
We use FLASH's N-body solver with $5\times10^{6}$ particles to realistically capture the response of the cluster potential to the induced gas sloshing.
We use particle density to refine our domain: when the number of particles in a block ($16^3$ cells) exceeds 1750, the block is refined, and conversely when the number of particles in a block falls below 1500, the block will be de-refined.
This allows us to achieve a resolution ranging from $\sim$~$9.76$\,kpc within a radius of $\sim$~$1$\,Mpc of the cluster core to $\sim$~$2.44$\,kpc within a radius of 0.22\,Mpc.
We run the simulations in a domain of 10\,Mpc$^3$ in size with diode (isolated) boundary conditions.
We allow the simulations to run for $\sim$~$10$\,Gyr with snapshots produced every 50\,Myr.
For simplicity we do not take account of cosmological expansion in the simulations, nor do we include radiative cooling or viscosity.

To automatically detect, and thus track, the SCFs in our simulations we use the SCF detection algorithm detailed in \citet{Vaezzadeh2022}, interpreting changes in the temperature profile of $>2\%$ over a radial range of 1.3\,kpc as SCFs, and discarding those fronts that have a temperature ratio between the start and end points of $<10\%$.
Once SCFs have been detected in this way, we relax the criterion that the temperature ratio across the front be $>10\%$ in order to carefully trace their evolution as far back in time as possible.
For our analysis of the simulations we use the Python based library, yt \citep{Turk2011}.

%*************
\subsection{Qualitative evolution of the Kick simulations}
\label{sec:qualitative}

\begin{figure*}
   \centering
   \subfigure{\includegraphics[width=0.24\textwidth]{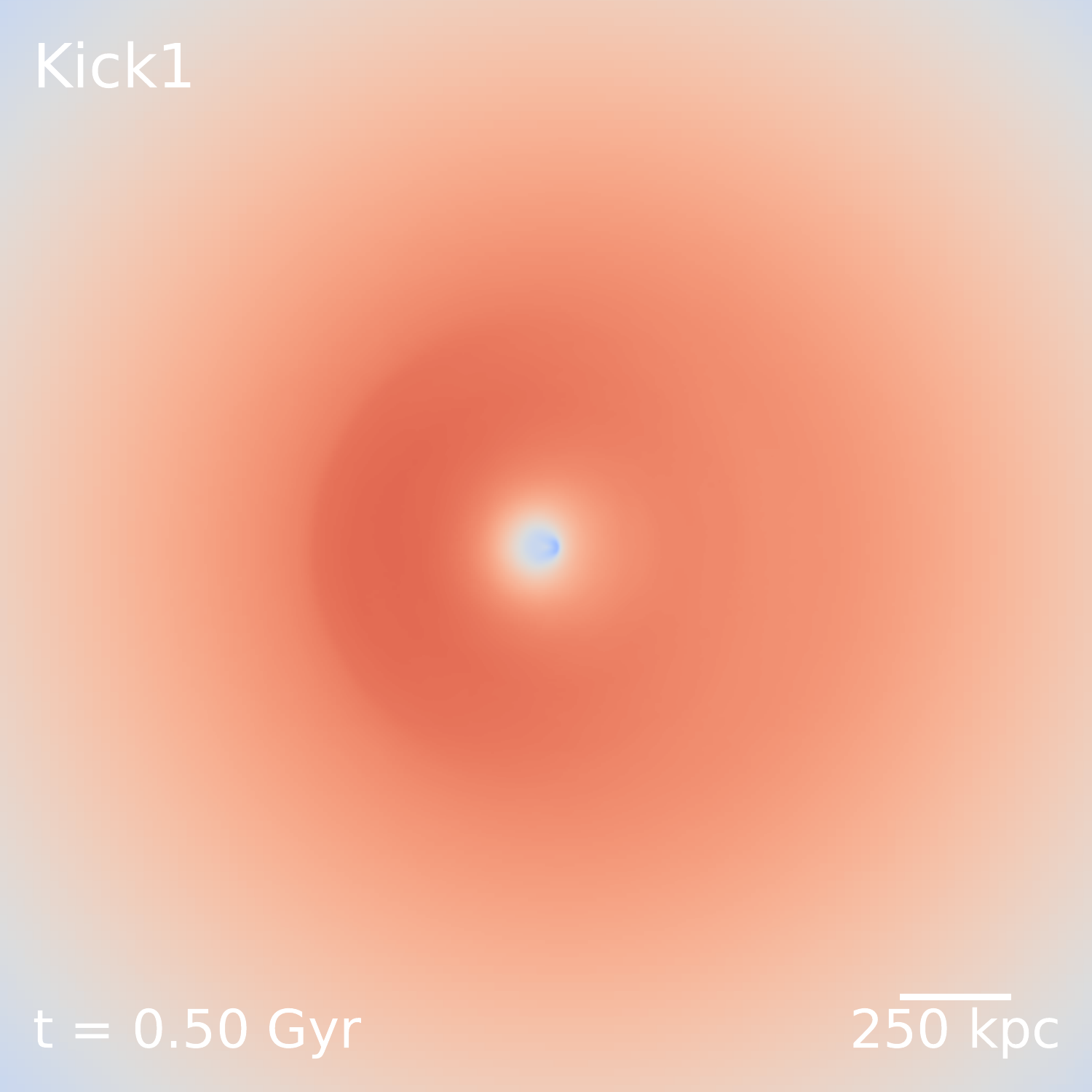}}\hspace{1mm}%
   \subfigure{\includegraphics[width=0.24\textwidth]{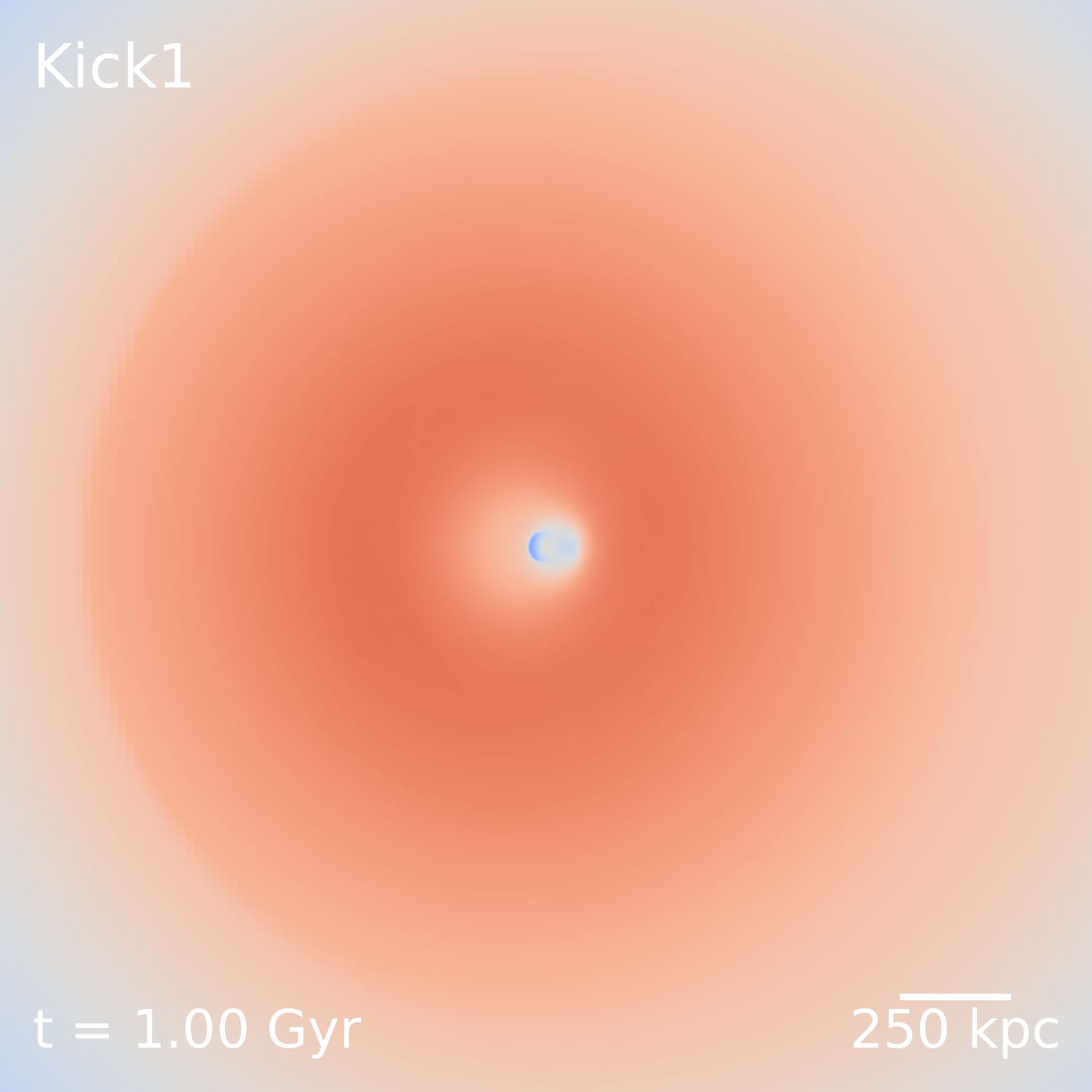}}\hspace{1mm}%
   \subfigure{\includegraphics[width=0.24\textwidth]{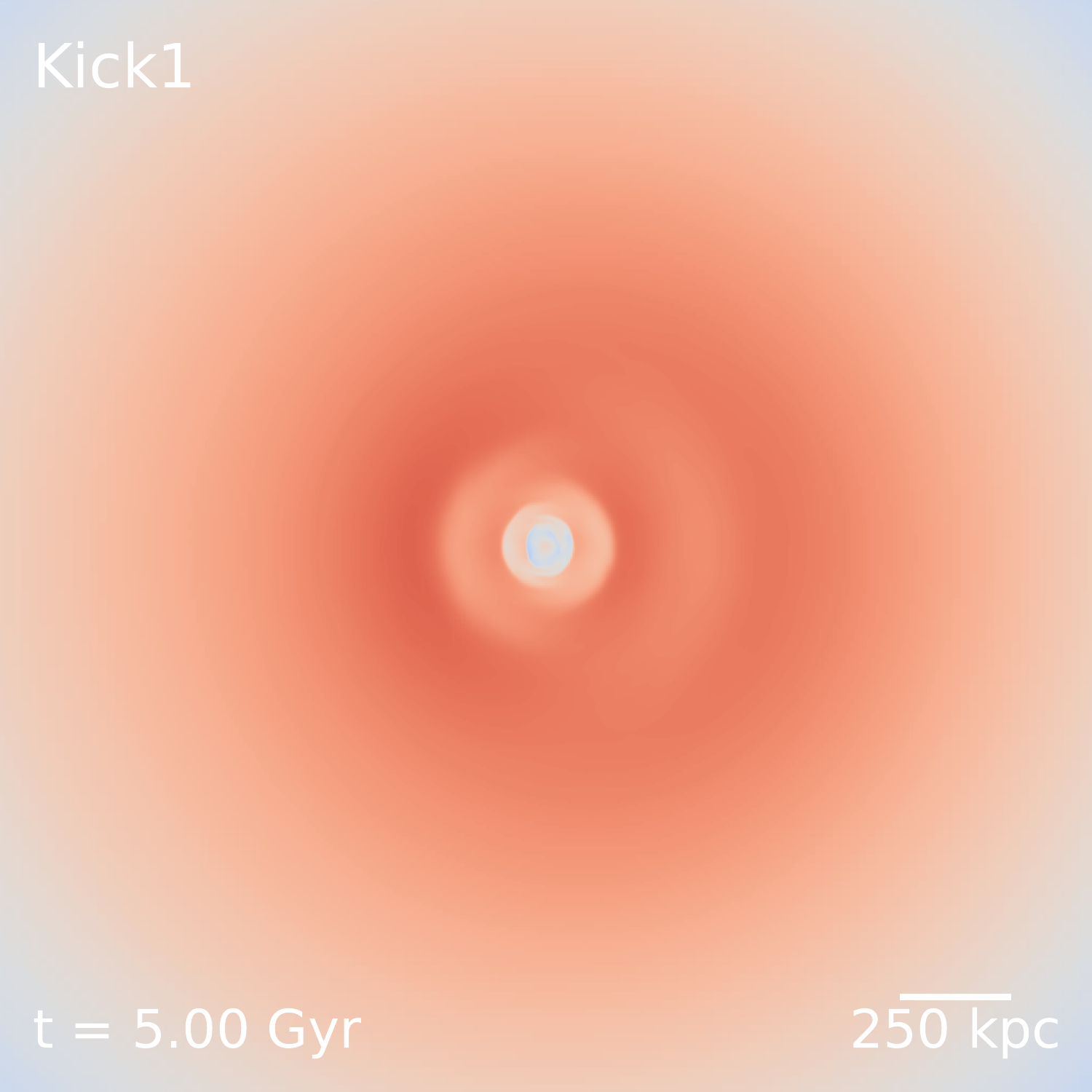}}\hspace{1mm}%
   \subfigure{\includegraphics[width=0.24\textwidth]{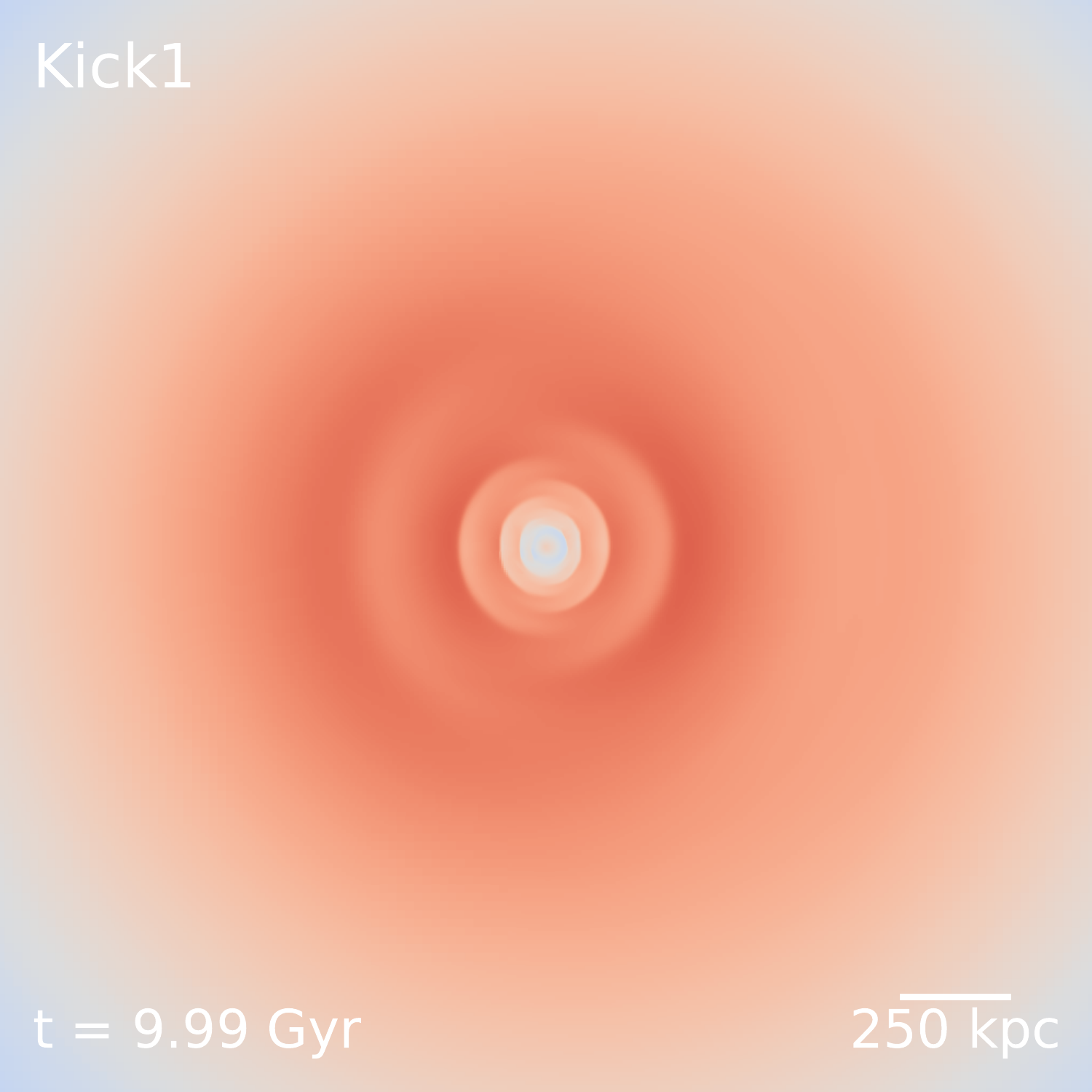}}\\[-3mm]% Adjust vertical spacing
   
   \subfigure{\includegraphics[width=0.24\textwidth]{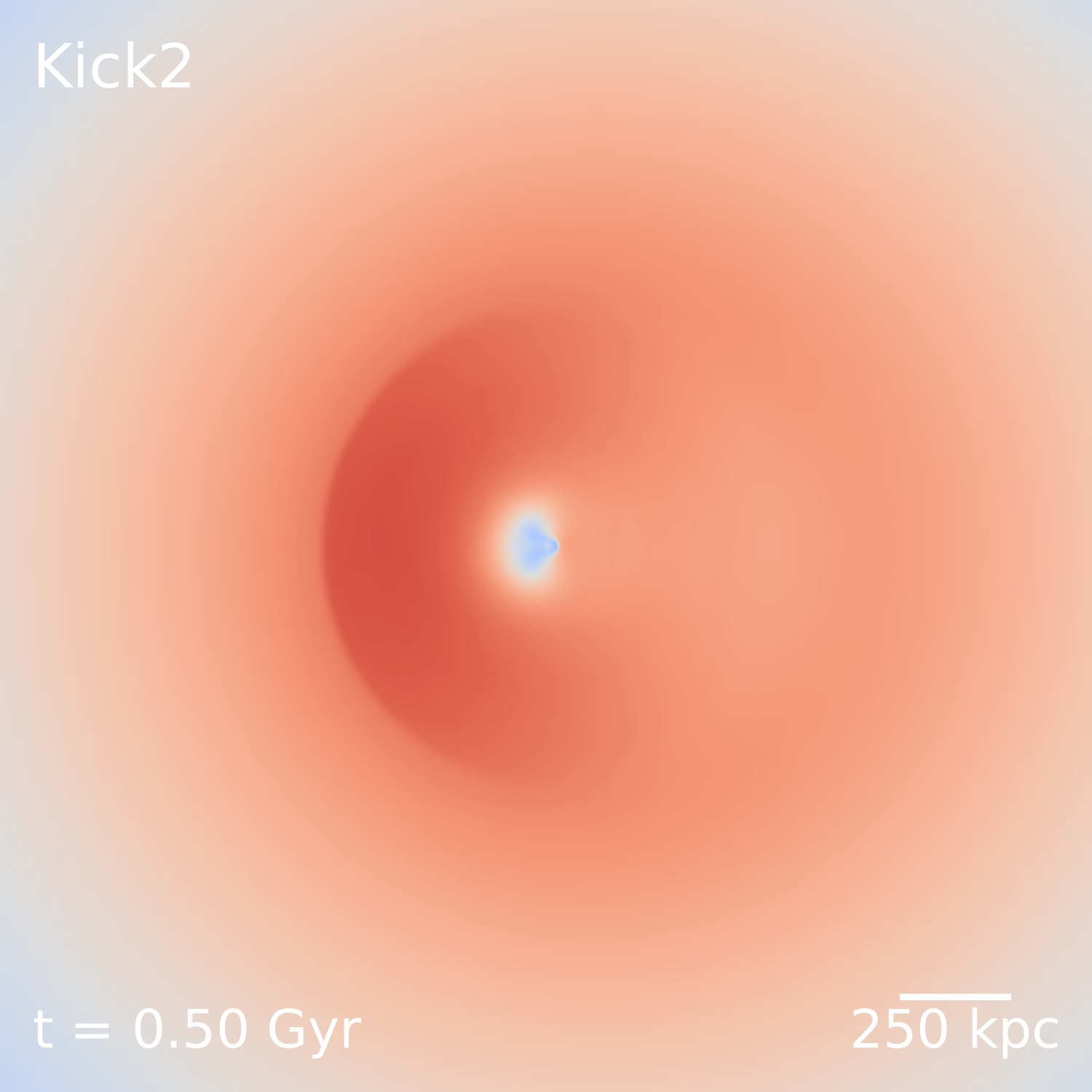}}\hspace{1mm}%
   \subfigure{\includegraphics[width=0.24\textwidth]{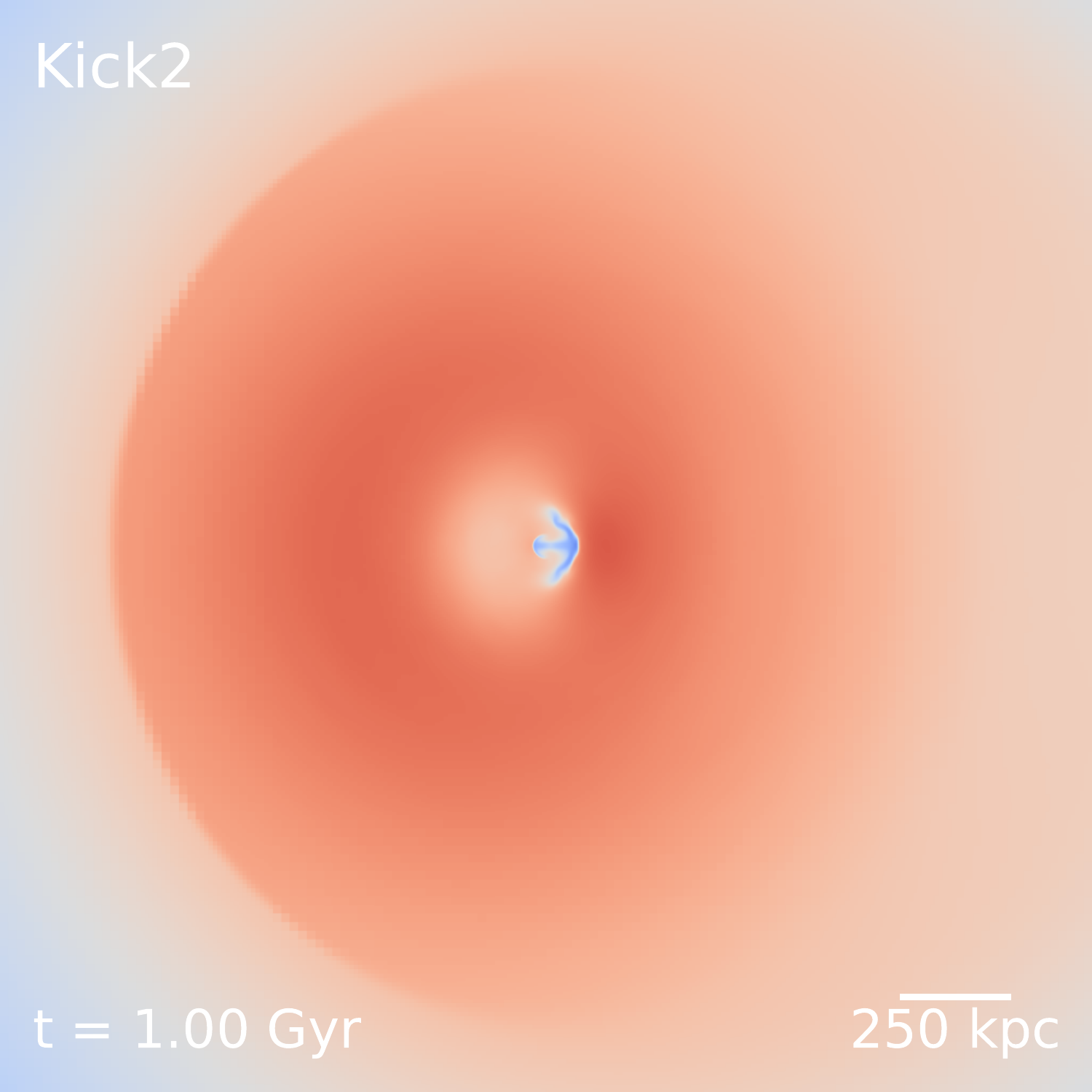}}\hspace{1mm}%
   \subfigure{\includegraphics[width=0.24\textwidth]{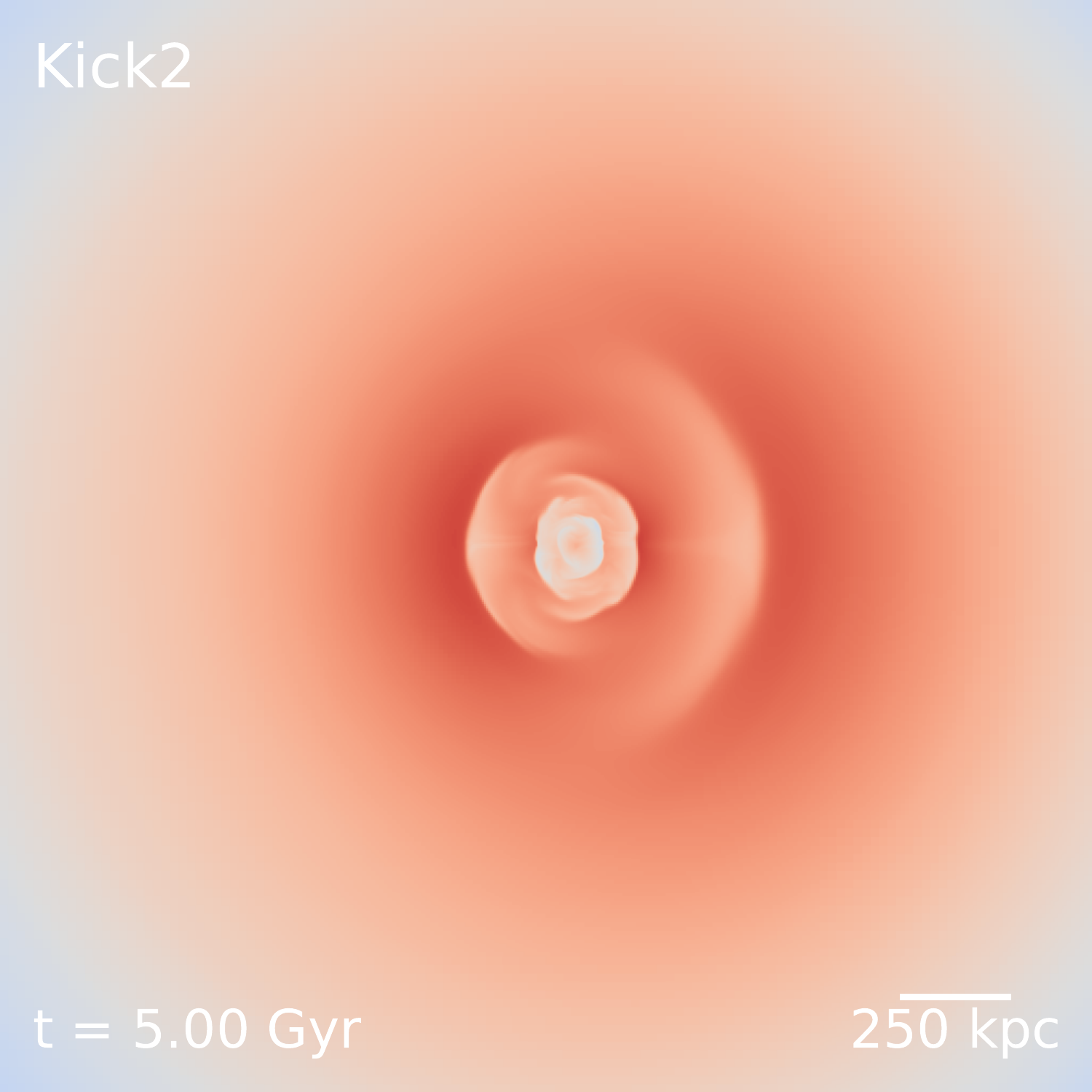}}\hspace{1mm}%
   \subfigure{\includegraphics[width=0.24\textwidth]{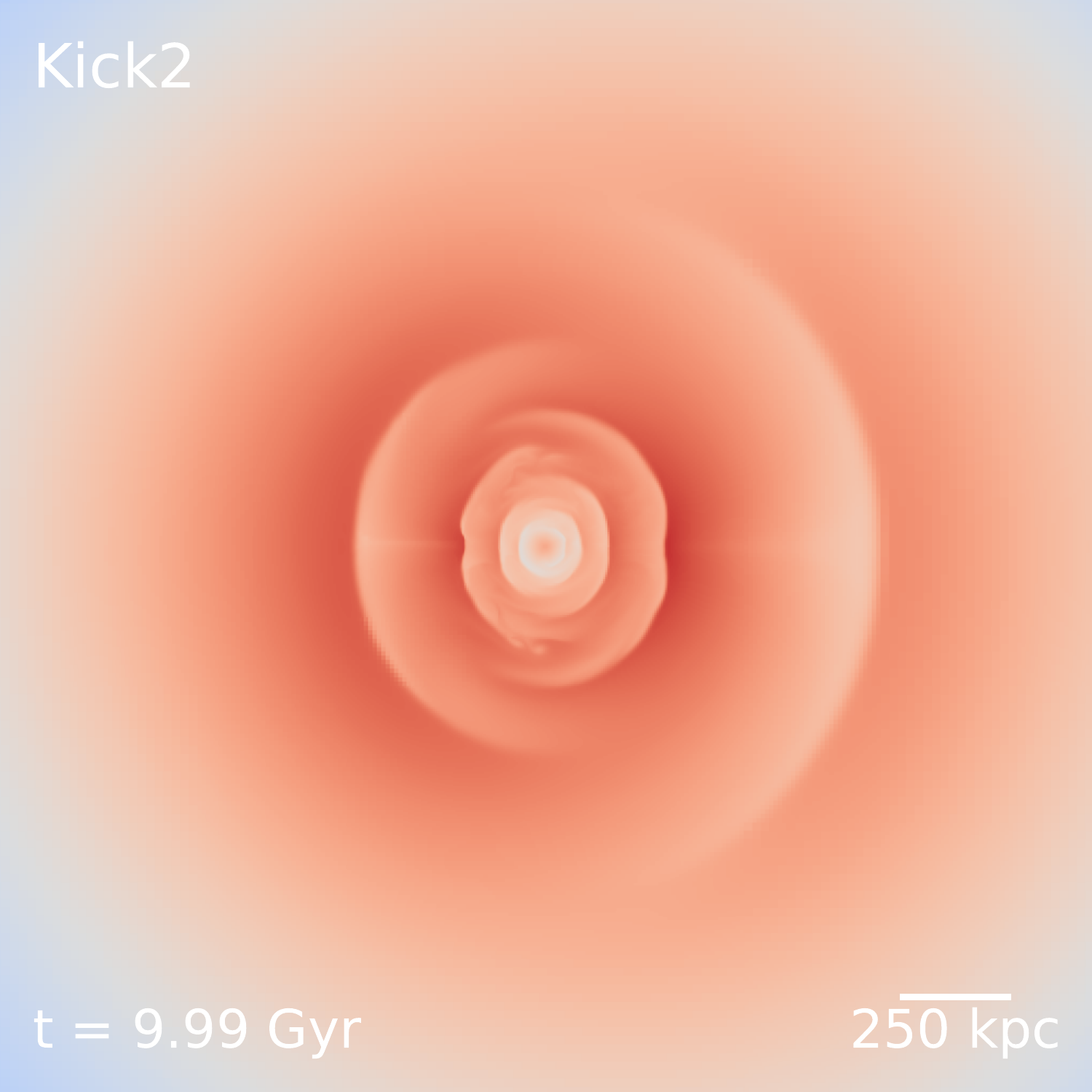}}\\[-3mm]% Adjust vertical spacing
   
   \subfigure{\includegraphics[width=0.24\textwidth]{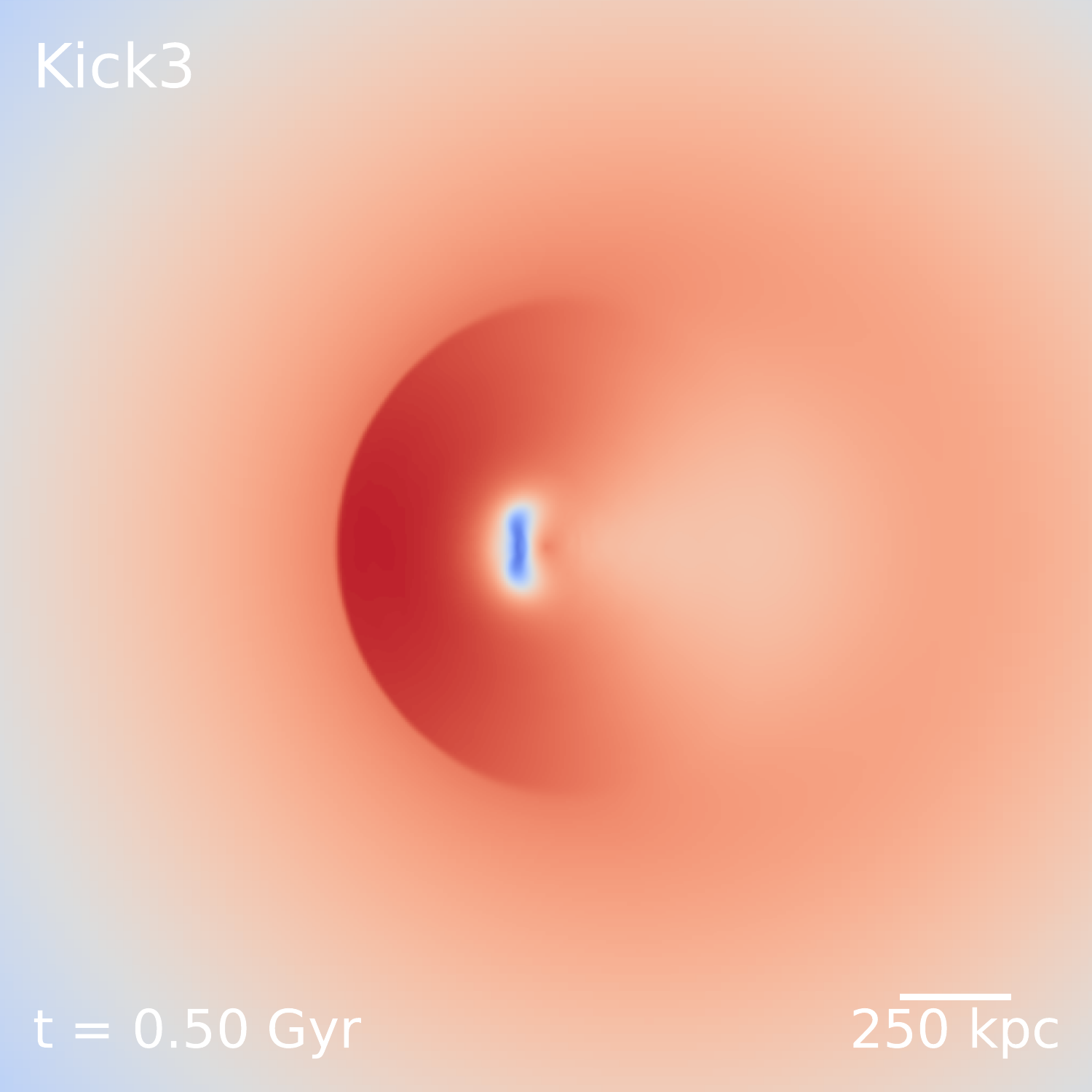}}\hspace{1mm}%
   \subfigure{\includegraphics[width=0.24\textwidth]{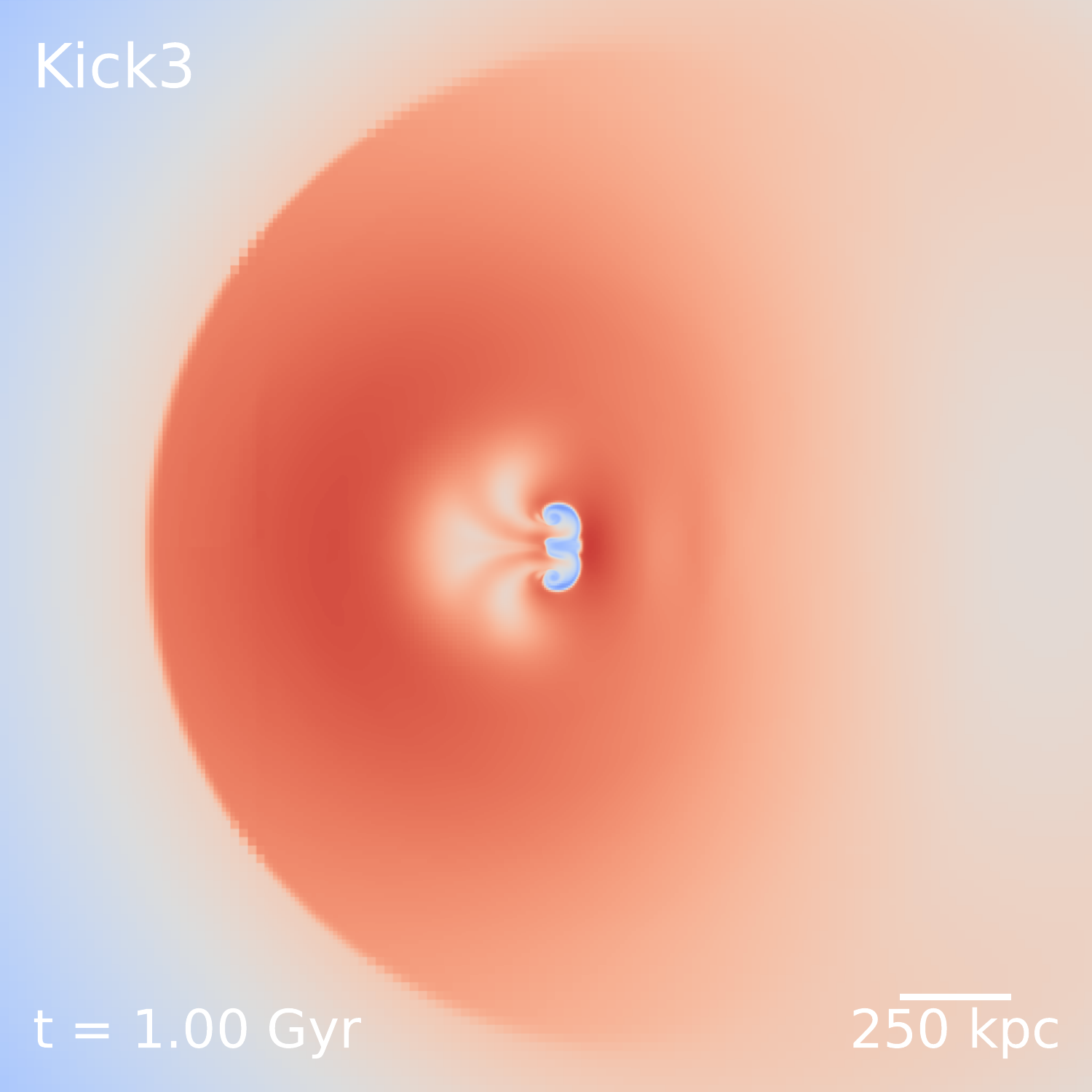}}\hspace{1mm}%
   \subfigure{\includegraphics[width=0.24\textwidth]{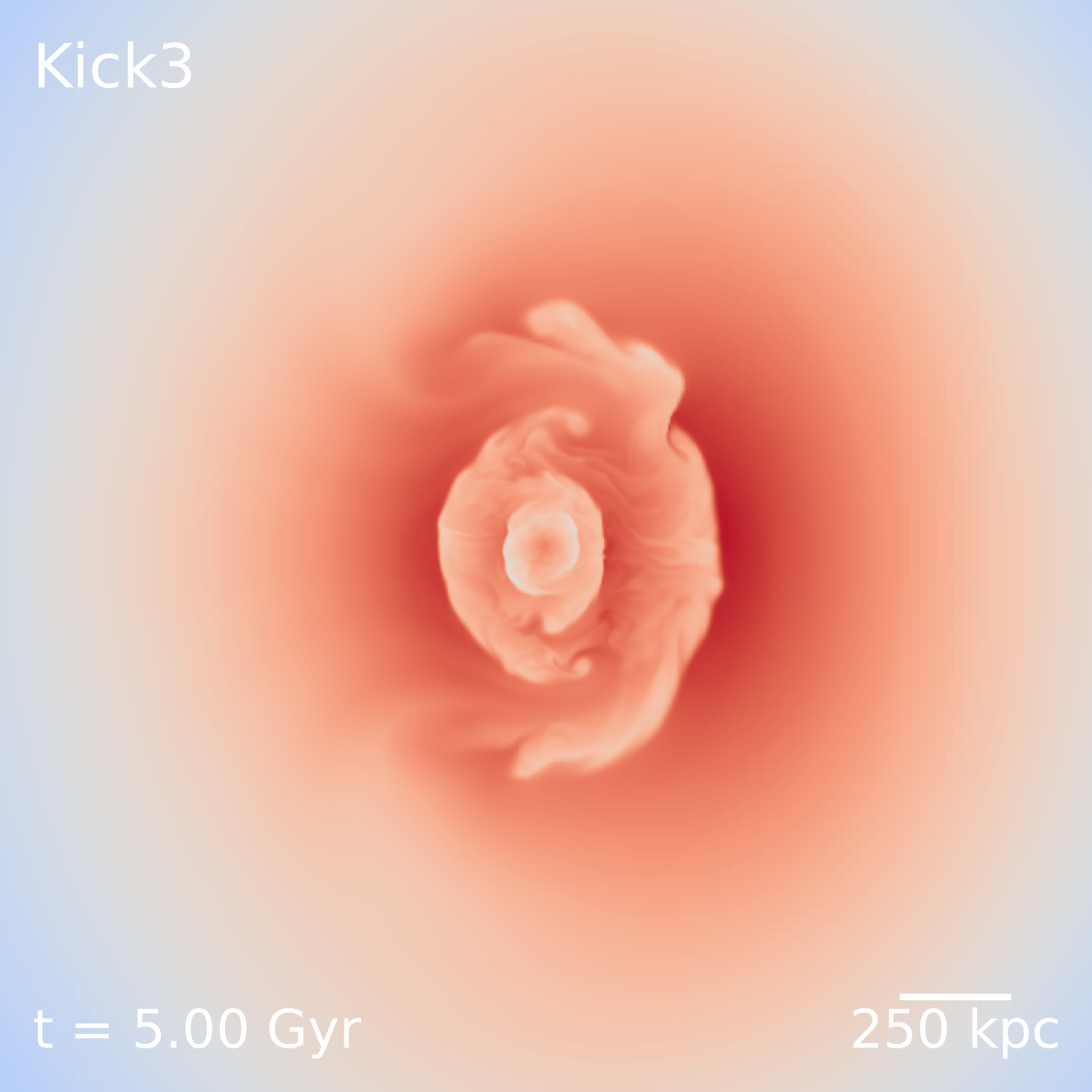}}\hspace{1mm}%
   \subfigure{\includegraphics[width=0.24\textwidth]{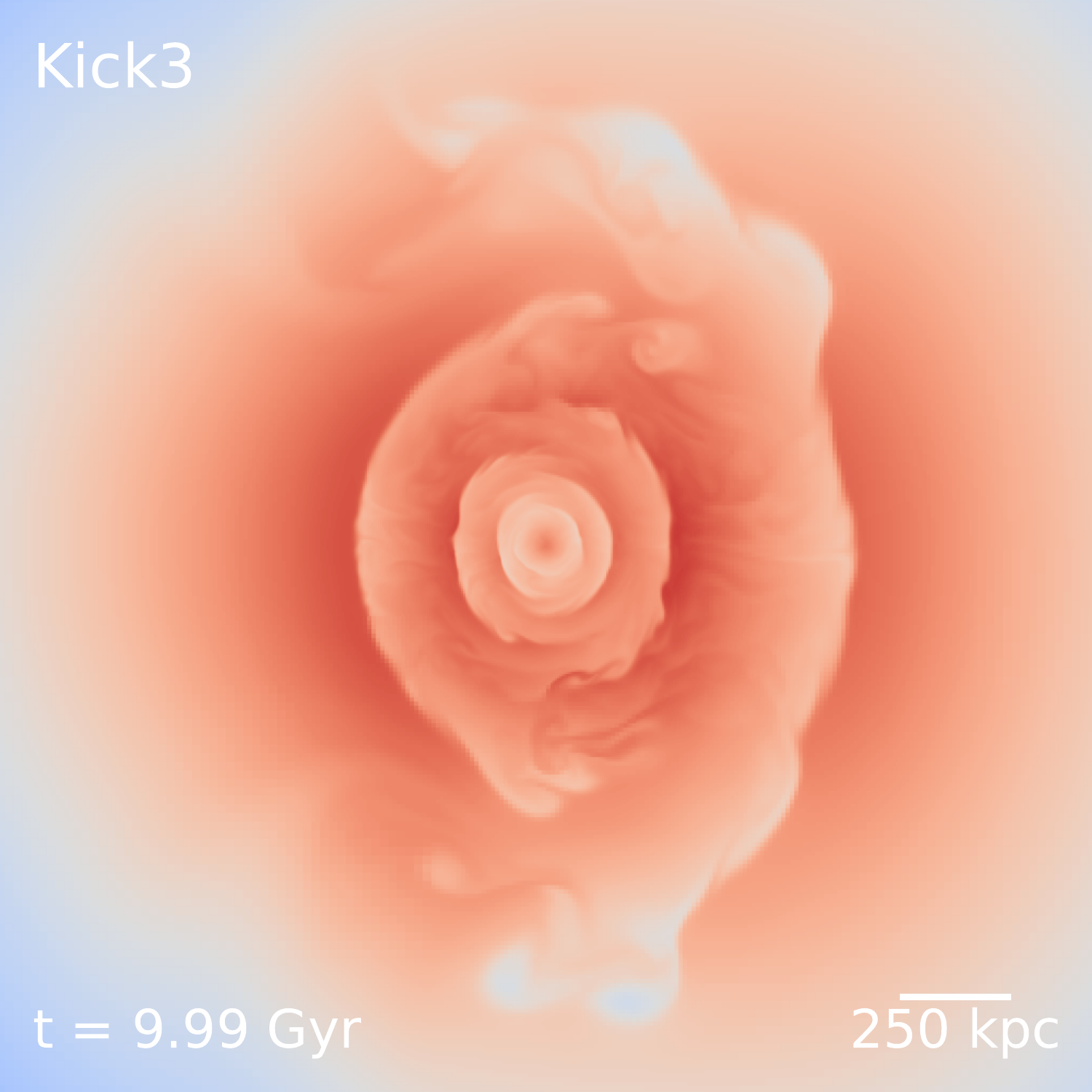}}\\[-3mm]% Adjust vertical spacing

   \subfigure{\includegraphics[width=0.24\textwidth]{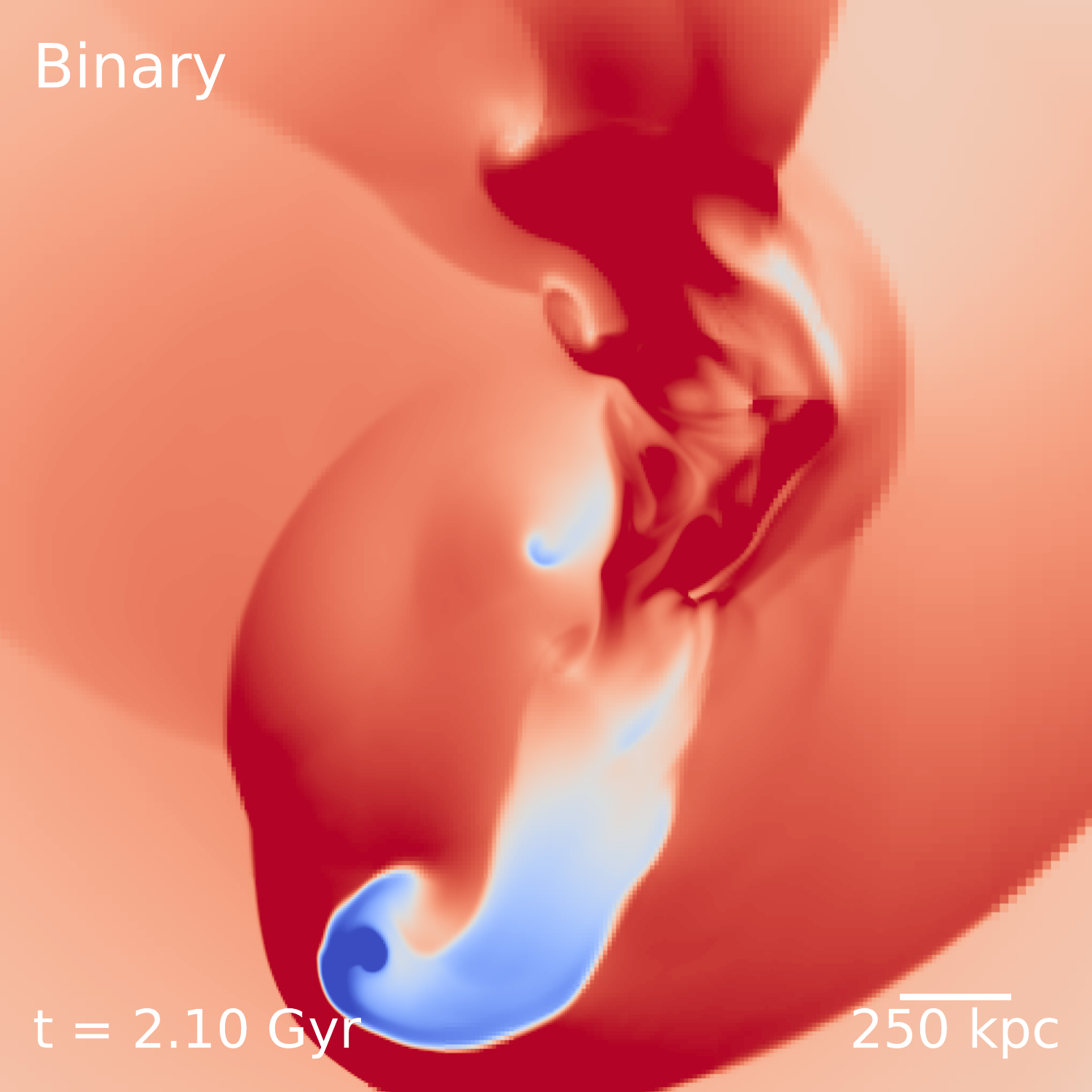}}\hspace{1mm}%
   \subfigure{\includegraphics[width=0.24\textwidth]{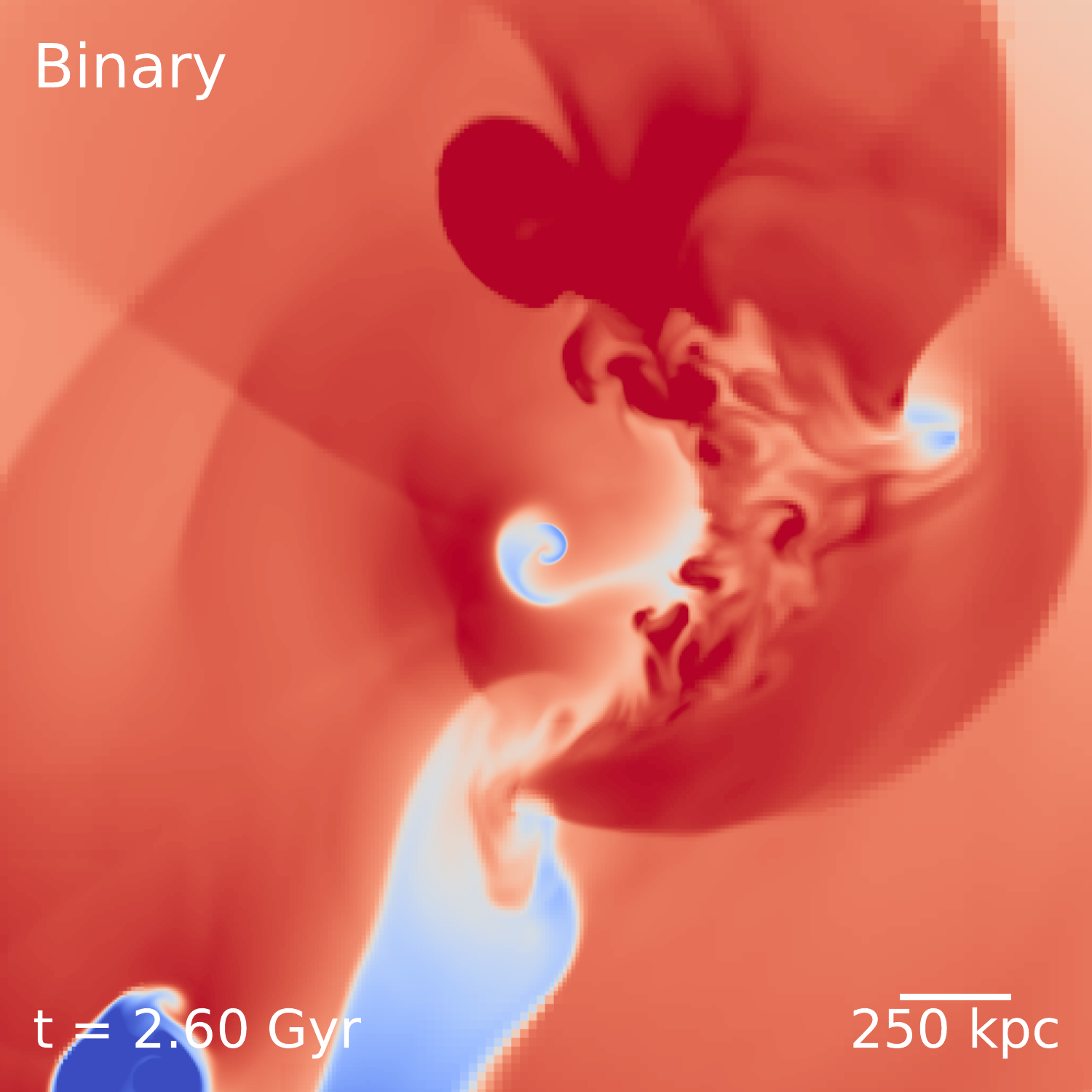}}\hspace{1mm}%
   \subfigure{\includegraphics[width=0.24\textwidth]{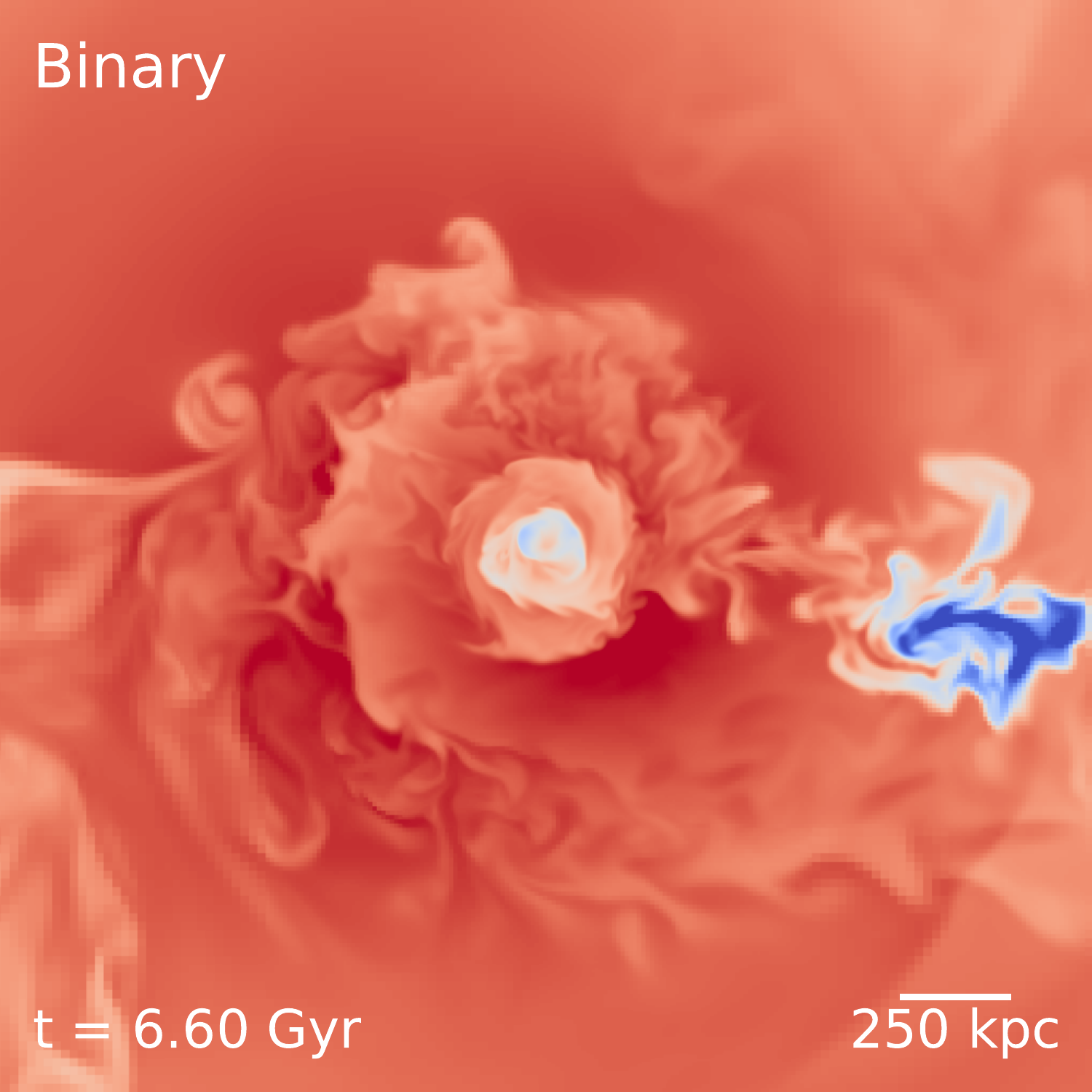}}\hspace{1mm}%
   \subfigure{\includegraphics[width=0.24\textwidth]{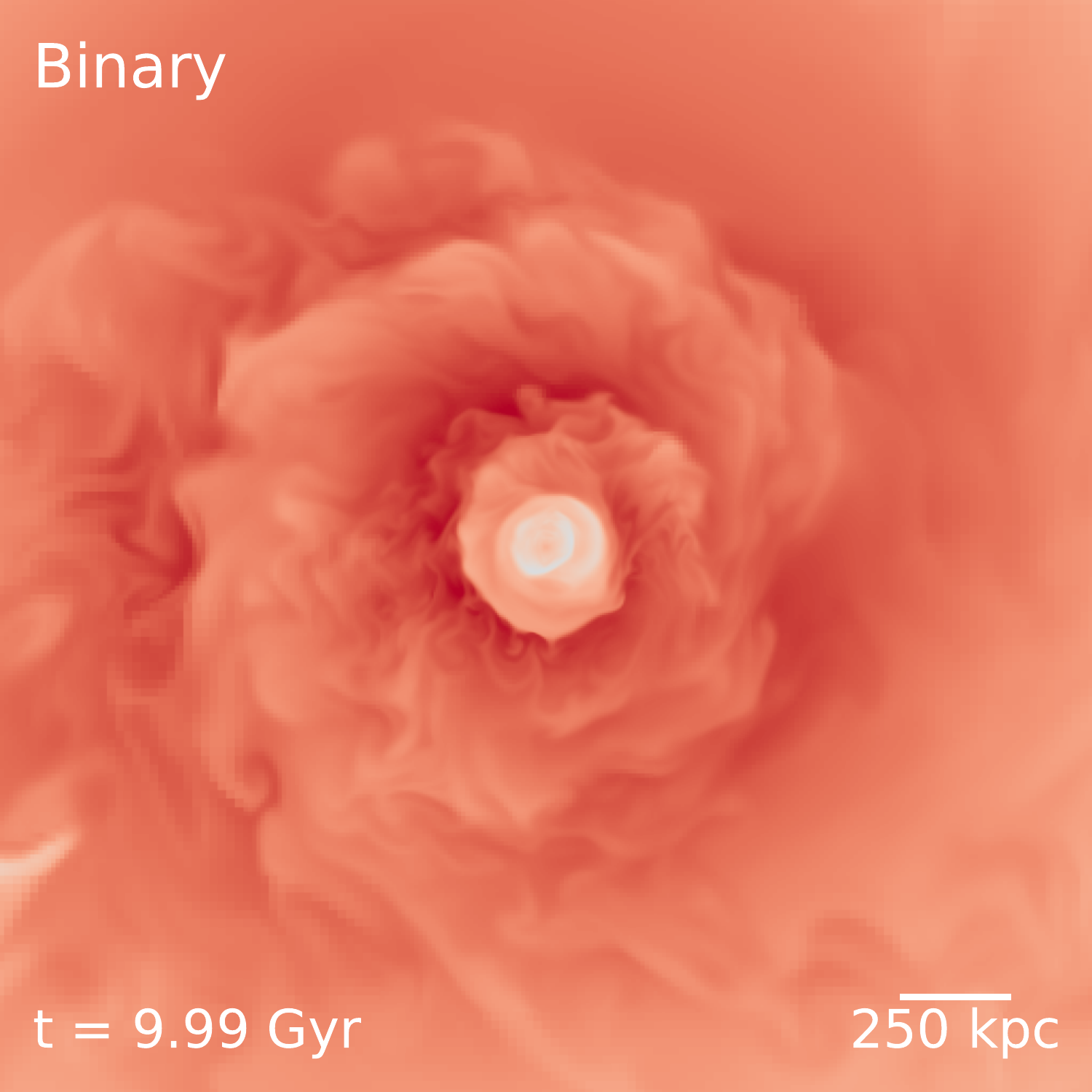}}\\[-0.5mm]% Adjust vertical spacing

   \subfigure{\includegraphics[width=0.4\textwidth]{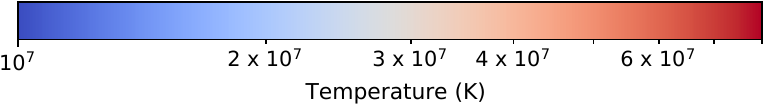}}
   \caption{Time series of temperature slices of the simulations centred on the minimum of gravitational potential.
   	Each row is for a different simulation, from top to bottom: Kick1, Kick2, Kick3, binary merger.
	The columns show the simulations at 0.5\,Gyr, 1.0\,Gyr, 5\,Gyr after the initial perturbation, and at maximum simulation time.
	As perturbation time, we take 0\,Gyr in the Kick simulations, and pericentre passage (1.6\,Gyr) in the binary merger.
	Each panel is 2.5\,Mpc on a side.
	A movie of the simulations can be found at \url{https://www.youtube.com/shorts/n9GBAzCs5T8}.}
   \label{fig:timeseries_all}
\end{figure*}

Figure~\ref{fig:timeseries_all} shows a series of snapshots for each of the Kick simulations (top 3 rows).
The sloshing process evolves very similarly despite the different kick strengths.
The motion of the ICM, after the kick along the $x$-axis, results in a sloshing pattern that is clearly orientated along the $x$-axis.
The arc-shaped sloshing fronts have an angular extent of approximately $180^{\circ}$.
Because the initial perturbation is perfectly axisymmetric (i.e.~contains no angular momentum), the sloshing pattern does not feature the characteristic `one-arm spiral' pattern so often seen in binary merger simulations.
The overall staggered pattern of SCFs is present with approximately the same size and staggering pattern in all three Kick simulations.
Indeed, at t$_{\mathrm{max}}$, corresponding SCFs across each simulation have positions within $\sim$~$20$\,kpc of one another.

In the case of Kick1, the first discontinuity emerges at $\sim$~$0.2$\,Gyr, travelling to the left (in the opposite direction to the kick).
However, this front is no longer detected by the detection algorithm beyond $\sim$~$100$\,kpc from the core, and is no longer visible by $\sim$~$0.5$\,Gyr, as seen in Figure~\ref{fig:timeseries_all}.
The second front emerges to the right of the core at $\sim$~$0.4$\,Gyr (as seen in the left hand column (0.5\,Gyr) of Figure~\ref{fig:timeseries_all}), and the third emerges at $\sim$~$0.7$\,Gyr to the left of the core.
The low initial gas velocity (in line with the small perturbation assumed by the toy model) has proven insufficient for these cold fronts to develop into true contact discontinuities, i.e.~they are `failed' fronts, and thus they cease to be detected by the detection algorithm at $\sim$~$1.85$\,Gyr, and $\sim$~$4.8$\,Gyr, respectively, though the structure is visible beyond these times.
Despite these fronts not having sufficient temperature jumps to be detected by the algorithm, we include them in subsequent analysis comparing toy model speed predictions to our simulations in order to maintain consistency between simulations.
At t$_{\mathrm{max}}$, Kick1 features eight SCFs (eleven if one includes the three `failed' outer SCFs).

Kick2, which has an initial gas velocity twice that of Kick1, proceeds in much the same way as Kick1 with the first front emerging at $\sim$~$0.15$\,Gyr, and ceasing its evolution at $\sim$~$0.4$\,Gyr.
The second front emerges to the right of the core at $\sim$~$0.45$\,Gyr, with subsequent fronts emerging on alternative sides of the core until $t_{\mathrm{max}}$.
A clear, staggered pattern of SCFs about the core is visible by t$_{\mathrm{max}}$, with little to no instability seen affecting the edges of the fronts.
There are nine SCFs visible at t$_{\mathrm{max}}$ (ten if one includes the initial failed front), with one less small radius front as compared with Kick1.

In the case of Kick3, the first cold front emerges at $\sim$~$0.1$\,Gyr, and travels left (opposite direction to the kick) but, as in the other simulations, it is no longer visible by $\sim$~$0.5$\,Gyr.
The second front then emerges at $\sim$~$0.6$\,Gyr to the right of the core.
From this point on, fronts continue to emerge on alternating sides of the core which continue to grow until t$_{\mathrm{max}}$, at which point the system features eight SCFs (nine including the initial failed front).
The large initial perturbation velocity (approximately half of the ambient sound speed of the cluster) causes sufficient disruption to the cluster that it can no longer be considered a cool-core cluster.
One would expect that this would stop SCFs emerging from the core of the cluster; however, fronts do continue to emerge, though there are fewer at small radii than in Kick1 and Kick2.
The large velocity also leads to clear instabilities along the edges of the cold fronts, with prominent Kelvin-Helmholtz instabilities (KHIs) visible.
It is interesting to note that these KHIs do not disrupt the SCFs sufficiently to hinder the fronts' growth and visibility.

%*************
\subsection{Tracking cold front position over time}
\label{sec:trackingCFs}

\begin{figure}
\centering
    \includegraphics[width=0.9\columnwidth]{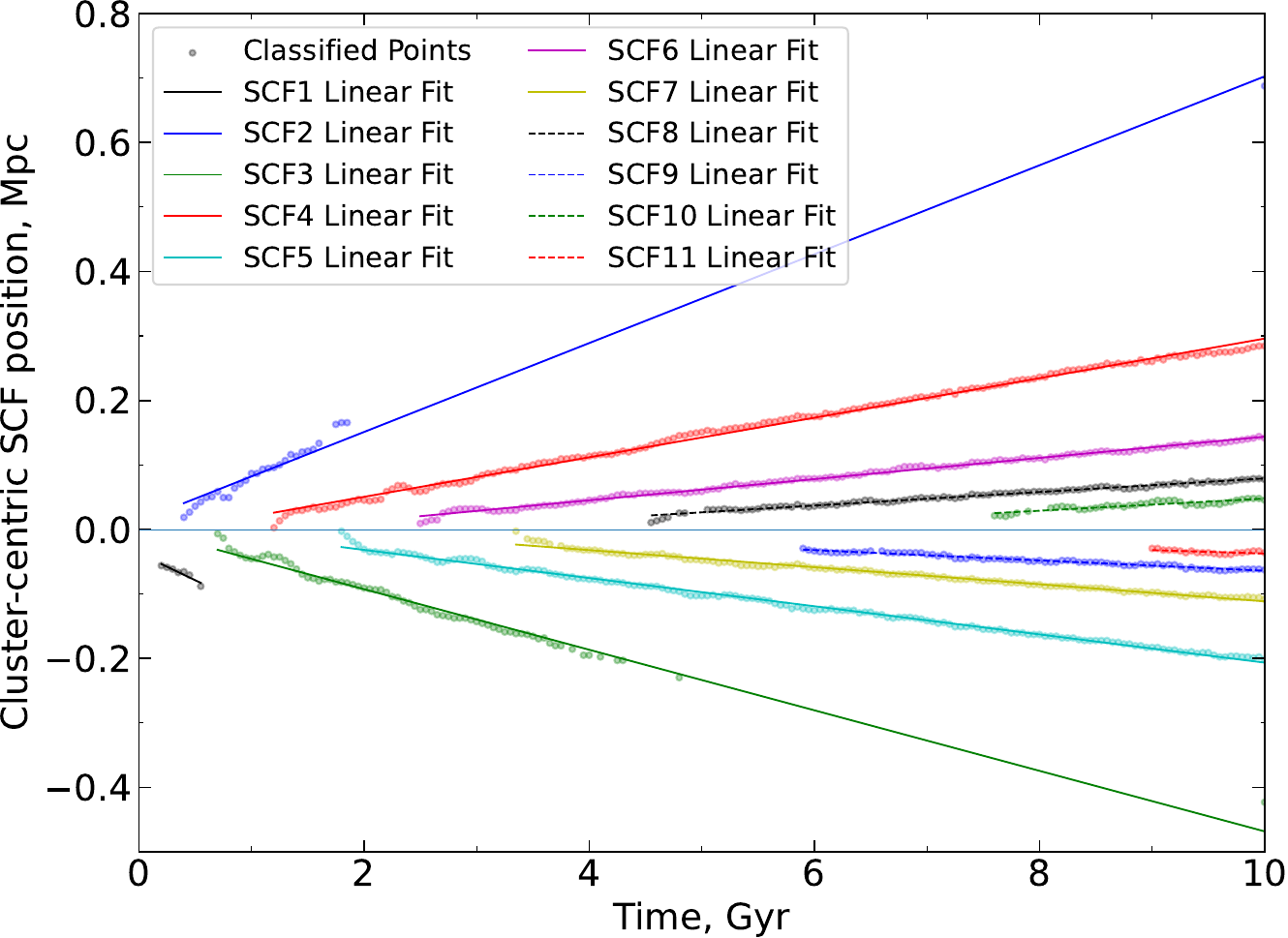}
    \includegraphics[width=0.9\columnwidth]{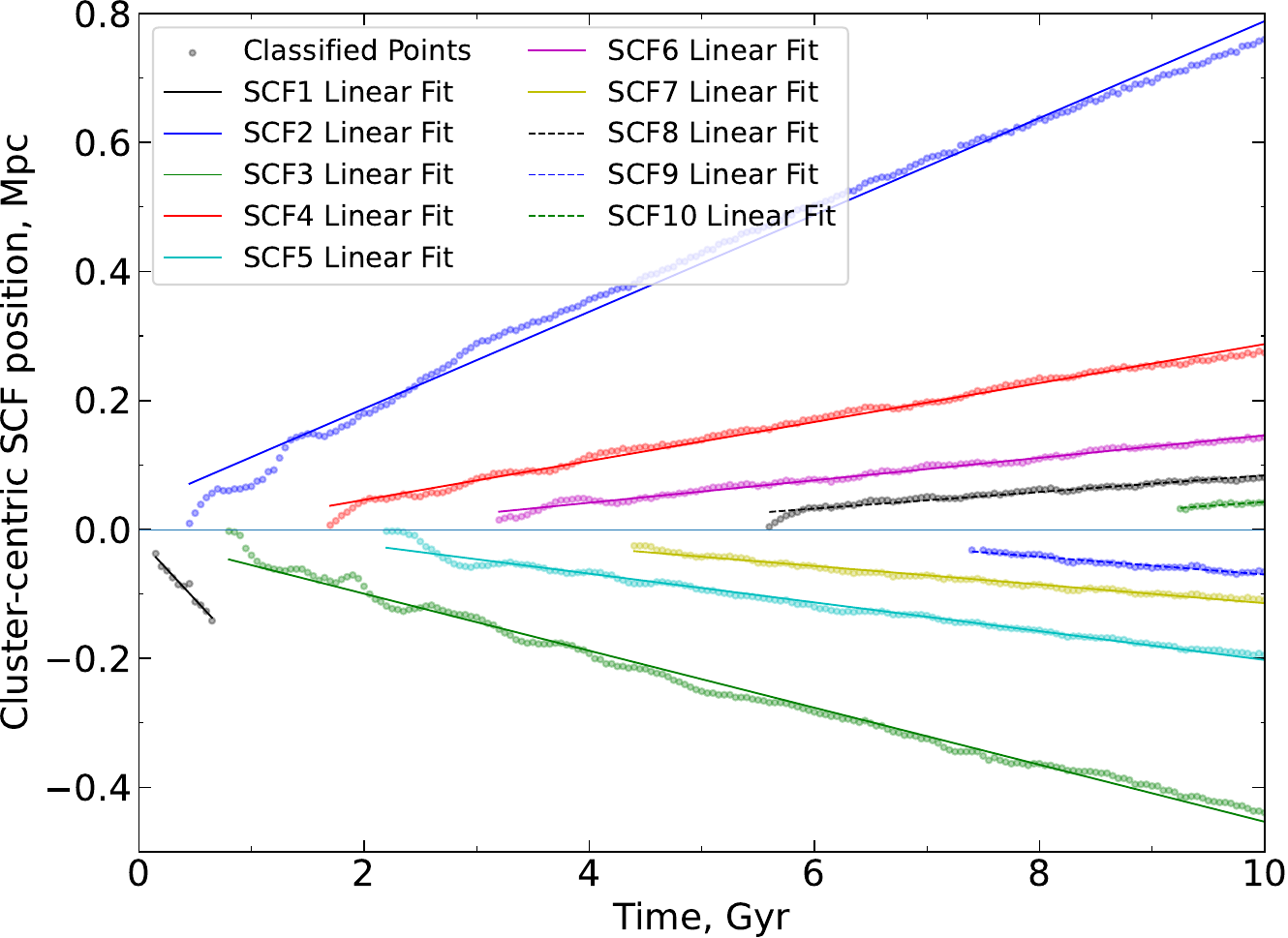}
    \includegraphics[width=0.9\columnwidth]{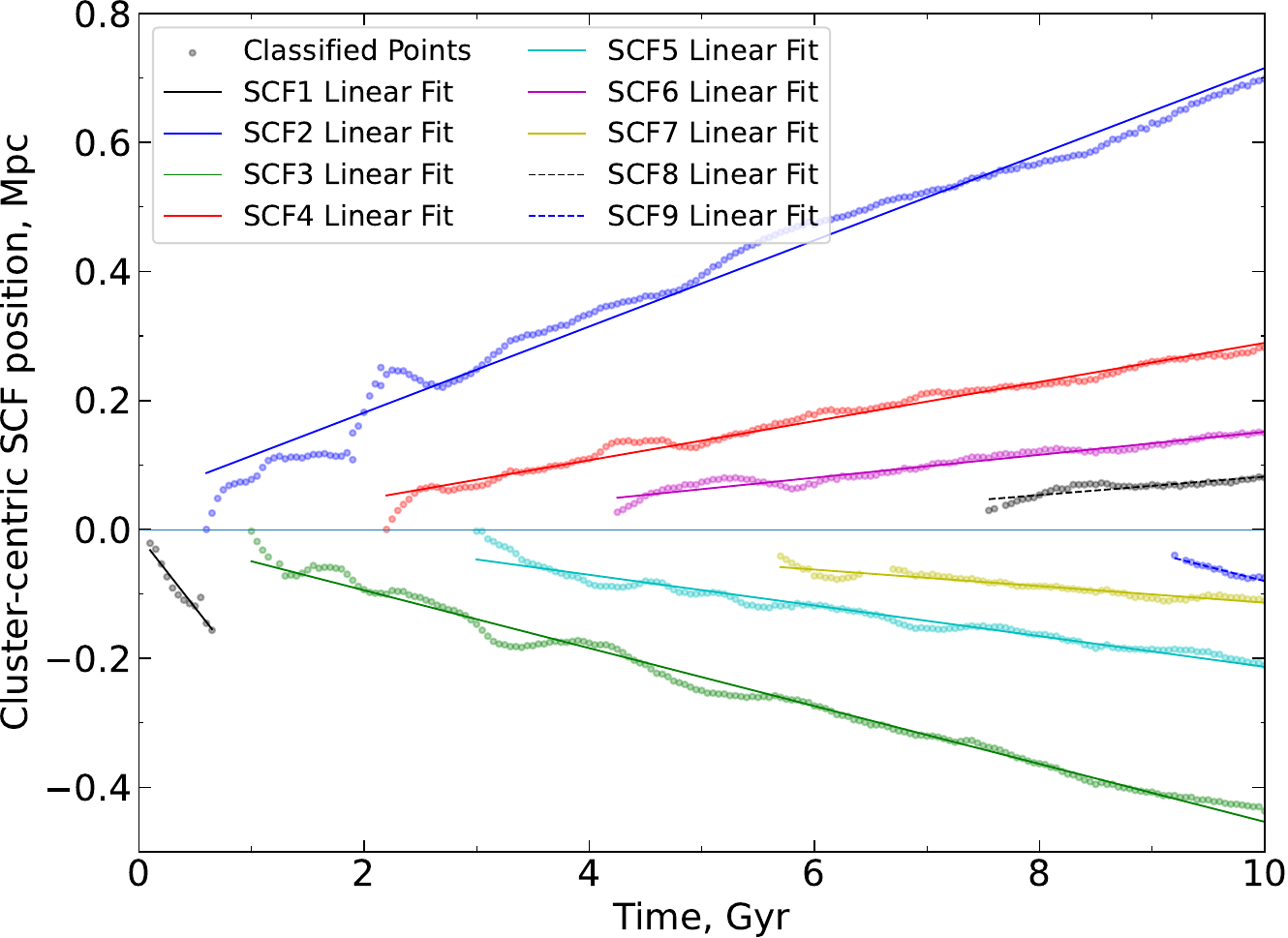}
\caption{Positions of detected cold fronts throughout each of the simulations Kick1, Kick2, and Kick3.
	The coloured dots represent distinct cold fronts.
	Solid lines of the same colour show linear fits to the cold front radii as a function of time.}
\label{fig:all_sims}
\end{figure}

Because the sloshing occurs along the $x$-axis (due to the perturbation being along the $x$-axis), we limit our analysis of SCF positions to their position along the $x$-axis.
The toy model predicts that the first front should emerge in the opposite direction to the perturbation, i.e.~the first CF should appear in the negative $x$-direction.
Figure~\ref{fig:all_sims} shows the positions of SCF detections throughout the duration of each simulation, with each SCF shown by a different colour.
As the first cold front in all three simulations fails at an early stage, we do not attempt to track its evolution beyond $\sim$~$0.5$\,Gyr.
As the next two outermost `failed' SCFs in Kick1 have ceased being detected automatically beyond $\sim$~$1.85$\,Gyr, and $\sim$~$4.8$\,Gyr, respectively, but are still visible by eye in temperature slices, we simply add their final positions manually by inspection of the slice images, such that Figure~\ref{fig:all_sims} captures their full motion.

SCFs move outwards with almost constant speeds, as predicted by the toy model.
The front speeds also decrease with each subsequent front that emerges, also in agreement with the toy model.
We then perform a linear regression on each individual SCF in Figure~\ref{fig:all_sims} within each simulation in order to extract speeds for each SCF.
Figure~\ref{fig:CF_speeds_comparison} summarises SCF speeds vs.~front number for the different simulation runs and the toy model.

According to the toy model, each SCF should emerge from the core of the cluster (i.e.~a radius of zero), and therefore a linear regression should be performed with the stipulation of a null $y$-intercept; however, leaving the $y$-intercepts as free parameters clearly gives the better fits to the overall motion.
The implied delayed emergence of inner SCFs is discussed below.

%*************
\subsection{Comparison of Kick simulations with toy model}
%*************
\subsubsection{Global features}
The toy model correctly predicts the outwards motion of sloshing fronts along the axis of initial perturbation, including the staggered pattern in the direction parallel and anti-parallel to the initial perturbation.
As predicted by the toy model, the sloshing fronts move outwards with approximately constant speed.

A substantial difference to the toy model is that the fronts `emerge' from the cluster centre one by one, each with a clear delay to the previous one of the order of 1\,Gyr, with the delay increasing with each subsequent front.
Figure~\ref{fig:CF_start_times} summarises these delay times for all simulations.
There is a clear trend in which the emergence time of each SCF is increasingly delayed as the kick strength is increased.
In contrast, the toy model predicts that the whole front system arises at once, and moves outwards in a self-similar fashion.
If this was the case, all graphs of front radii as a function of time should start from $r=0$ at $t=0$, but this is not the case in the simulations.

The toy model correctly predicts that lower perturbation speeds lead to `failed' outer cold fronts, i.e.~features that are temperature gradients in the correct direction, but are not discontinuities.
For example, in the case of Kick1 (the weakest perturbation case) the three outermost fronts are `failed' fronts, whereas the fronts further inwards are discontinuities.
This behaviour is predicted by the toy model qualitatively, and even quantitatively: according to Equation~\ref{eq_vkick_strength}, for the characteristic speed $u=134$\,km/s, and kick speed $100$\,km/s, only fronts with $n\ge 2\times 134/100 = 2.68$ should be discontinuities, which therefore predicts one more successful front than is observed.
Performing this calculation for Kick2 results in a prediction of one failed front, which agrees with the simulation.
In the case of Kick3, the toy model predicts no failed fronts, but in the simulation the first front fails.

%*************
\subsubsection{Cold front speeds and positions}
\begin{figure}
    \includegraphics[width=0.94\columnwidth]{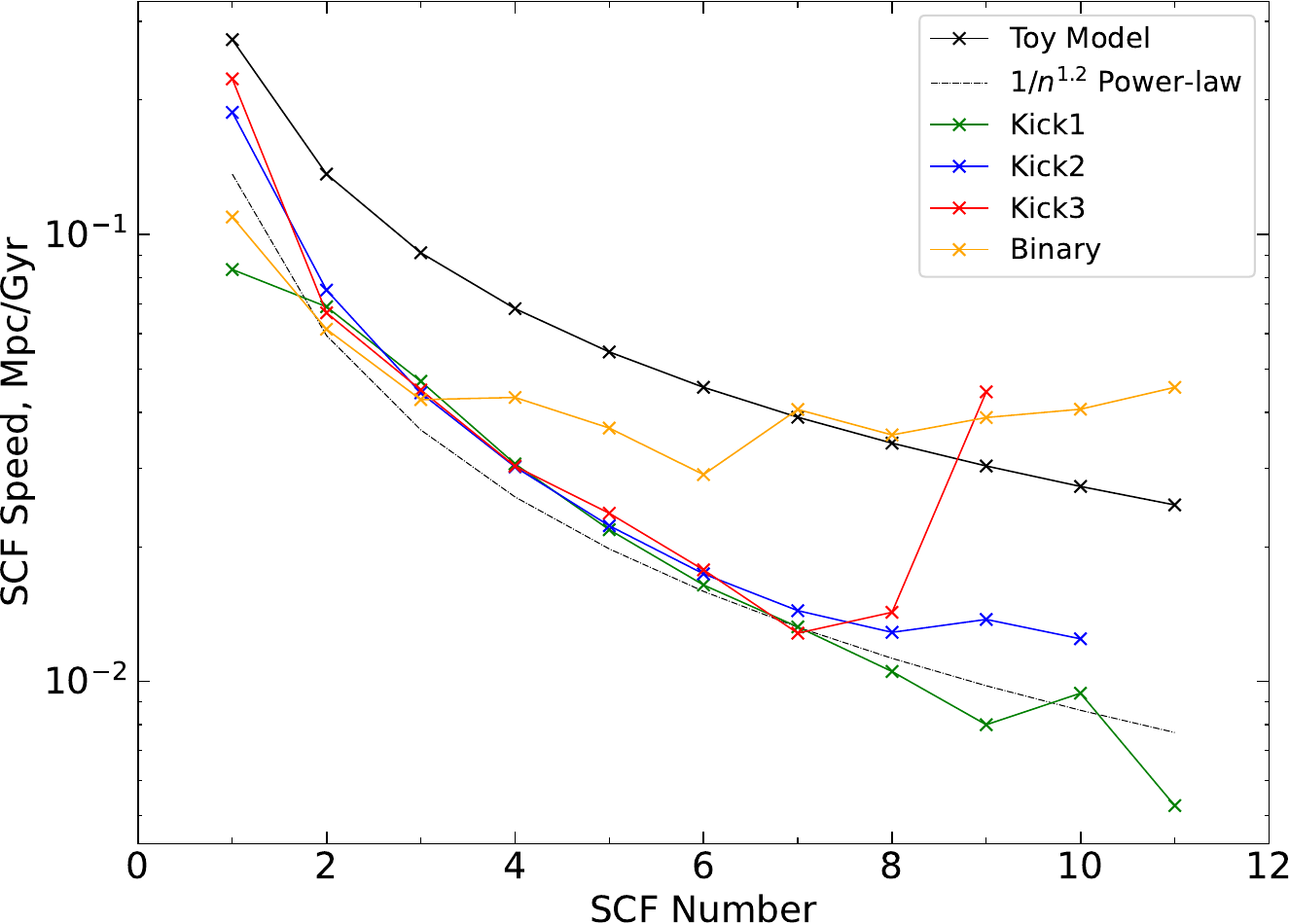}
\caption{Lineplots of the linear best-fit SCF speeds from each simulation, and the speeds predicted by the toy model as a function of SCF number.
	The solid black line shows the toy model predictions for each SCF's speed; the solid coloured lines show the linear fit SCF speeds from each of the simulations; and the dash-dotted black line shows a power-law of $1/n^{1.2}$.}
\label{fig:CF_speeds_comparison}
\end{figure}

\begin{figure}
\centering
    \includegraphics[width=0.9\columnwidth]{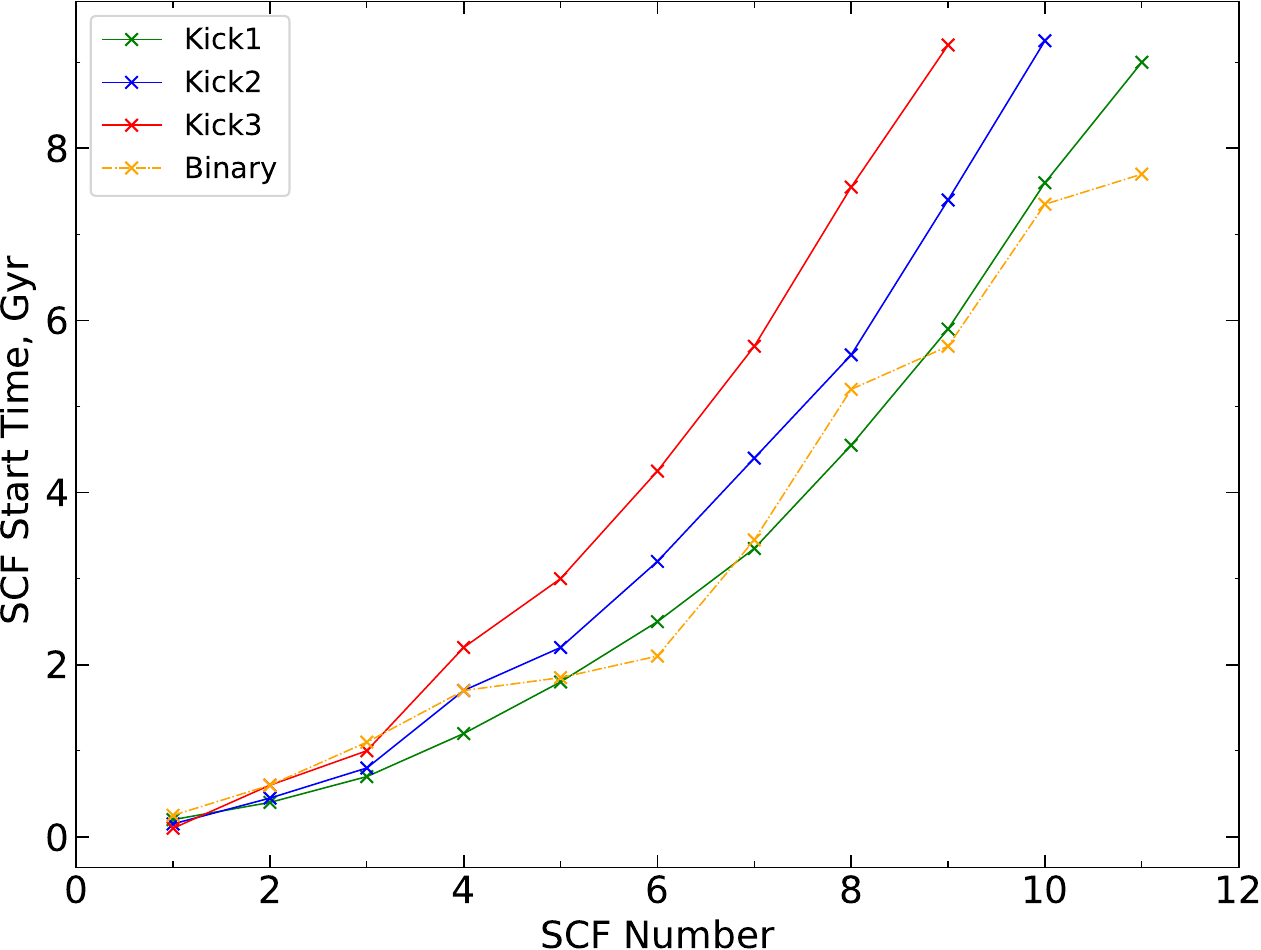}
\caption{The observed start times of each SCF in each simulation.
	The times are aligned by the perturbation time (i.e.~1.6\,Gyr in the binary merger, and 0\,Gyr in the Kick simulations).}
\label{fig:CF_start_times}
\end{figure}

Figure~\ref{fig:P_slosh} shows the sloshing timescale as calculated from the Brunt-Väisälä frequency as a function of radius for our model cluster.
Clearly the sloshing timescale is monotonically increasing with radius in an approximately linear fashion within $r_{200}$.
Departures from linearity in the function will cause deviations from the simple model outlined here.
We do not see any SCFs travel beyond $r_{200}$, and so we limit our analysis to within this radius, where the linear approximation is appropriate.

\begin{figure}
\centering
    \includegraphics[width=0.9\columnwidth]{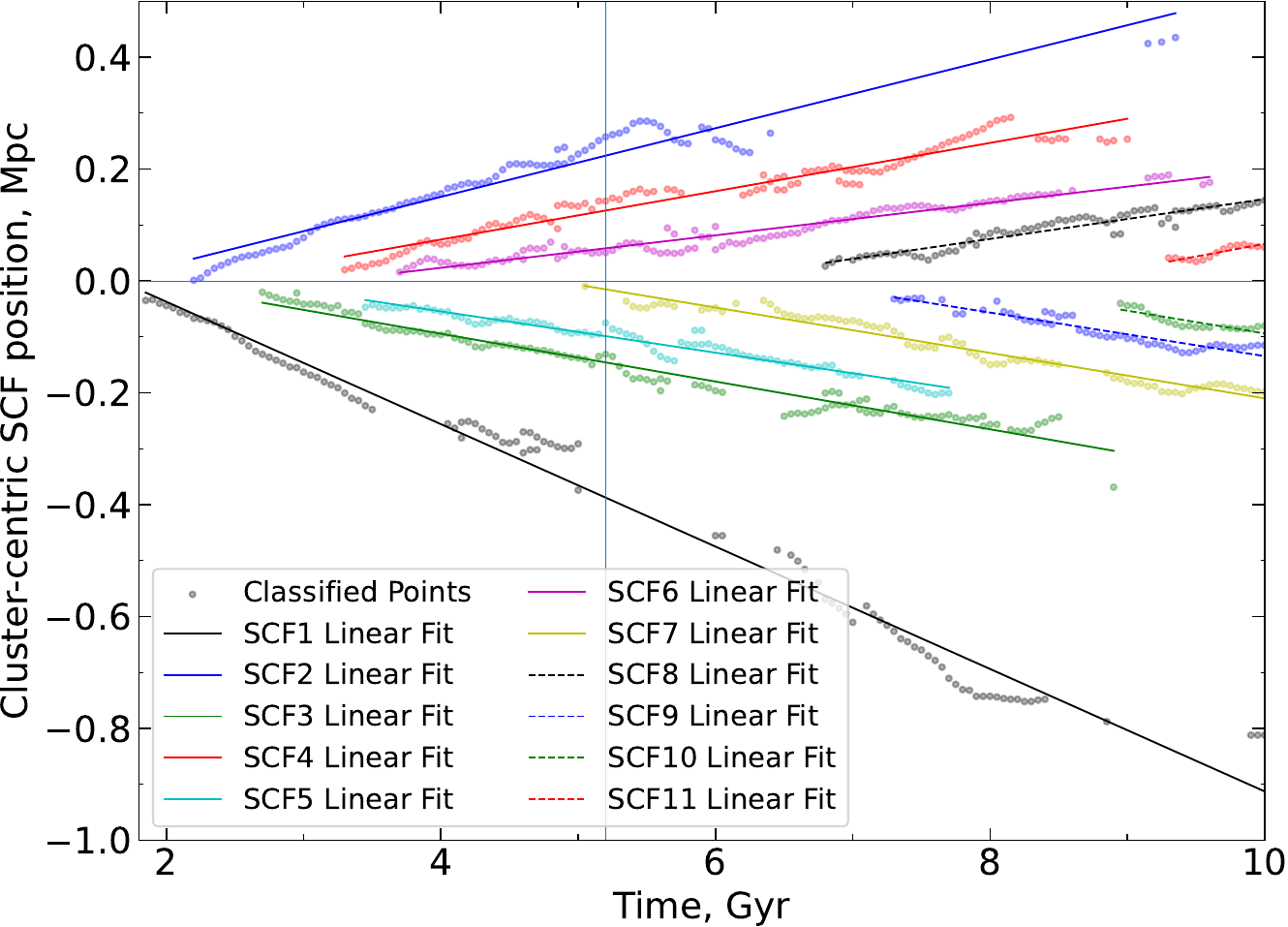}
\caption{Positions of detected cold fronts, along the $x$-axis, throughout the binary merger simulation.
	The data points begin at 1.6\,Gyr, which is the time of first pericentre; the vertical blue line at 5.2\,Gyr shows the time of second pericentre.
	The coloured dots represent different tracked SCFs, with the correspondingly coloured solid lines representing linear fits to those tracked SCFs.}
\label{fig:Binary_posns}
\end{figure}

From Equation~\ref{eq_u_final}, we obtain the characteristic sloshing front speed in this cluster of $u\approx 134$\,km/s.
We can then scale this speed via Equation~\ref{eq_oscillatorSpeeds} to predict a unique speed for each of the SCFs that emerge during the course of the simulations.
As explained above, the counter, $n$, counts the sloshing fronts from the outermost one inwards.
Odd fronts arise in the direction opposite to the kick directions, i.e.~along the $-x$-direction, and even-numbered fronts arise in $+x$-direction.
The solid black line in Figure~\ref{fig:CF_speeds_comparison} shows the CF speed as a function of front number as predicted by the toy model.
The coloured lines show the CF speeds derived for the Kick simulations from Figure~\ref{fig:all_sims} and for the binary merger from Figure~\ref{fig:Binary_posns}.

Figure~\ref{fig:CF_speeds_comparison} reveals that, overall, the CF speeds in the Kick simulations are about a factor of 2-3 below the predicted value, and that the front speed depends on front number in a similar power law fashion as predicted (power $-1.2$ instead of $-1$).

Differences between the Kick simulations occur for the outermost front, and for fronts beyond number 7.
The outermost front in Kick1 was a `failed' front, and moves slower than the one in the other Kick simulations.
The speed of the outermost CF between Kick2 and Kick3 agrees well.
The deviation from the predicted pattern at higher CF numbers could arise because the sloshing process changes the inner entropy profile of the cluster, and thus the conditions of the initial state, assumed by the toy model throughout, are not true anymore.
This behaviour is increased by the fact that inner cold fronts indeed arise with a delay.
More deviations from the initial entropy profile are expected to arise with increasing time, and with increasing perturbation strength --- both are seen in Figure~\ref{fig:CF_speeds_comparison}.

Figure~\ref{fig:CF_start_times} summarises the emergence delay times of SCFs in the different simulations.
This behaviour is not predicted by the toy model.
These delay times were derived by taking the earliest time at which each SCF could be seen, and aligning these times relative to the perturbation time (0\,Gyr in the Kick simulations, and 1.6\,Gyr in the binary merger).
Given these unpredicted delay times, the toy model alone will not be able to correctly predict the positions of sloshing fronts.
While the front speeds could be calibrated, the toy model does not predict the delay in the inner cold fronts emerging.
Given their slow speed, this delay has a big impact on the actual CF position at a given time.

%*************
\subsection{Comparison to idealised binary cluster merger}
\label{sec:binary_sec}

In order to test our model in a more realistic ICM sloshing scenario, we compare our toy model's predictions to a binary cluster merger with mass ratio, $R=1:10$, in which the primary cluster is the same cluster as in the other simulations presented in this paper, with the exception of it having a `WHIM'-like atmosphere beyond 2.17\,Mpc.
This `WHIM' is a uniform gas background with density, temperature, and pressure values of $1.03\times10^{-29}\,$g/cm$^3$, 1.70\,keV, and $2.85\times10^{-14}$\,erg/cm$^3$, respectively.
The `WHIM' has the effect and purpose of pre-truncating the atmosphere of the infaller to avoid it carrying too much gas into the primary.
The infalling subcluster has a mass of $5 \times 10^{13} M_{\odot}$, with an $r_{200}$ of 777\,kpc, and a particle resolution of $5 \times 10^5$ particles.
The subcluster is initialised at 2.45\,Mpc north of the primary (the sum of the respective $r_{200}$ radii), with a radial velocity of $-950$\,km/s, and a tangential velocity of $450$\,km/s such that the subcluster will pass to the right of the primary with a large pericentre distance, and thus deliver a `kick' along the same axis as in the simulations presented in the previous section.

%*************
\subsubsection{Considerations regarding the nature of the perturbation}
It is important to note that the key difference between the Kick simulations presented in the previous section, and the binary merger, with regard to the toy model, is the initial perturbation that the cluster receives.
The perturbation is continuous, and non-constant in space and time in the binary merger, i.e.~the infaller is already perturbing the primary cluster when the simulation begins, and continues to do so as it moves through the primary.
Furthermore, the binary merger introduces angular momentum into the host cluster.
The perturbation may be a mix of extended kicks and offsets in the primary's ICM.
It is also not obvious which direction of perturbation matters most -- on first approach, the perturber attracts the primary's ICM towards it, but then pushes and pulls the primary's ICM somewhat along its orbit after it passed a given location.
This has important implications for the application of the toy model, as the toy model assumes a single `kick' or offset to a row of oscillating gas parcels which then oscillate independently.
In the binary merger case, the perturber first moves approximately parallel to the $y$-axis, but its second passage through the cluster occurs more in a diagonal direction from the $-x$, $-y$ quadrant to the $+x$, $+y$ quadrant.

We pointed out that the perturbation in a minor merger is continuous.
However, for the sake of comparison, we simplify this scenario to the often-invoked model of sloshing in which the first pericentric passage is the key moment of perturbation.
As the infaller passes to the right of the primary in our binary merger simulation (therefore pulling the primary to the right at pericentre time), the `kick' is to the right, and therefore sloshing will occur along the $x$-axis.
We note that it is significantly more difficult to trace each SCF for its full evolution than in the case of the highly idealising sloshing simulations.
Due to the highly `messy' nature of the SCFs' evolution in the binary merger, automatic tracking is more difficult, and is therefore augmented by manual tracking of the SCFs.
Due to the angular momentum imparted by the infaller, the primary's ICM is swirling at the same time that it is sloshing, and as such fronts that emerge along the $x$-axis to a given side rotate around to the other side of the cluster in some cases.
Once the coherent SCF points have been identified, the same procedure of linear regression is applied as described in Section~\ref{sec:trackingCFs}.

%*************
\subsubsection{Comparison of binary merger with Kick simulations and toy model}
Figure~\ref{fig:Binary_posns} shows the CF positions as a function of time along the $x$-direction.
Again, to each CF we fit a linear position-time function to determine the speed of each CF.
Figure~\ref{fig:CF_speeds_comparison} compares the CF speeds from the binary merger to the toy model prediction and the Kick simulations.
Figure~\ref{fig:CF_start_times} compares the delay times of the emergence of the SCFs in the binary merger to the delay times for the Kick simulations.

Despite the significantly more complex perturbation in the binary merger, sloshing fronts also arise in a staggered pattern, and move outwards with about constant speed.
Similar to the Kick simulations, the inner sloshing fronts emerge with a delay, which is not predicted by the toy model.
The dependence of delay time on SCF number is approximately linear, and of a similar order to that seen in Kick3 for early SCFs, and Kick1 for later SCFs.

The SCF speeds (Figure~\ref{fig:CF_speeds_comparison}) are again within a factor of 2 of the toy model prediction.
For SCF numbers of 5 and below, the SCF speeds show a similar dependence on SCF number as the ones in the Kick simulations.
For outer SCFs, there is an approximate agreement with the $1/n$ dependence, although there is significantly more scatter around that trend.
From CF 6 onwards, the SCF speeds increase again, even becoming faster than predicted by the toy model.

It is worthwhile noting that, despite the slightly lower overall evolution time in the binary merger (the `kick' at pericentre occurs at 1.6\,Gyr as opposed to 0\,Gyr), there is an equal number of SCFs in the binary at $t_{\mathrm{max}}$ to in Kick1.
The ICM bulk motions in the primary cluster triggered by the merger are about 600\,km/s, which is comparable with the kick speed of simulation Kick3.
Thus, the perturbation in this binary merger is not a particularly weak perturbation compared to the Kick simulations, especially given that it is a continuous perturbation.

%*********************************************************
\section{Discussion} \label{sec_discussion}
%*********************************************************
We presented a simple toy model for the sloshing process in galaxy clusters.
The toy model assumes that sloshing arises as the ICM parcels in an initially hydrostatic cluster start oscillating around their equilibrium radius with their local Brunt-Väisälä (BV) period after an initial perturbation.
We showed in Equation~\ref{eq_T_BV_linear} that the BV period in galaxy clusters can be approximated as a linear function of radius.
The variation of the BV period with radius leads to a characteristic pattern of density enhancements in the ICM that can be linked to sloshing cold fronts.
These enhancements travel outwards with about constant speed for a range of initial perturbations (instantaneous kick, instantaneous offset, instantaneous local perturbation that travels through the cluster supersonically).
We compared the toy model's prediction in detail to hydrodynamic+N-body simulations of both sloshing initiated by an initial kick to the ICM, and sloshing caused by a binary minor cluster merger.

%*************
\subsection{Successes of the toy model}
\label{sec:discussion_success}

The toy model correctly predicts several key qualitative characteristics of the sloshing front system in a galaxy cluster:
\begin{itemize}
    \item Sloshing fronts arise from the centre of the cluster, and move outwards, i.e.~at any time there is an outermost SCF.
    	The physical way of numbering SCFs should start with this outermost front.
    \item Sloshing fronts form a staggered pattern, i.e.~appear in an alternating fashion on opposite sides of the cluster core.
    \item Sloshing fronts move outwards with about constant speeds.
    \item For the outer few SCFs, the front speed decreases with front number, $n$, approximately in proportion to $1/n$.
    \item Weak perturbations can lead the outermost fronts to be `failed' sloshing fronts, i.e.~they are not discontinuities but have sloshing cold front characteristics in every other aspect.
\end{itemize}
The toy model quantitatively predicts SCF speeds within a factor of 2 to 3.
The characteristic front speed is 14\% of the cluster's sound speed (Equation~\ref{eq_u_final}).
Thus, the toy model predicts subsonic motion of sloshing fronts.
These features have been seen in hydrodynamic simulations (e.g.~\citealt{Ascasibar2006,ZuHone2011}), including the feature of `failed' SCFs \citep{Roediger2011}, and are reproduced in our idealised Kick simulations presented here as well as our binary merger simulation.

Given that sloshing fronts are a wave phenomenon, we cannot expect sloshing to transport matter over large distances.
For example, we do not expect sloshing to transport low entropy gas from cluster centres to the outskirts.

%*************
\subsection{Limits of the toy model}
\label{sec:discussion_limits}

The first major deviation between the toy model and the hydrodynamic simulations is the delay time of inner, i.e.~later, cold fronts.
In simulations with either an initial instantaneous, kick-like perturbation as well as perturbation by a classic binary minor merger, our analysis showed that the sloshing fronts emerge from the cluster centre with a substantial delay time that grows with cold front number.
This means that the position of a given cold front depends not only on its speed, but also on its emergence time.
The toy model does not predict this emergence delay, but predicts that the whole sloshing front system emerges together, and grows in a self-similar pattern.
Further investigations are required to uncover the origin of the emergence delay of later CFs, and why the toy model still predicts reasonable cold front speeds despite this mismatch.

Secondly, the toy model over-predicts the speed of the sloshing fronts by a factor of 2-3.
This is related to the fact that that the toy model considers only radial oscillation modes.
In the full treatment (Nulsen et al., in prep.), the oscillation frequency depends on the angle $\theta$ between the wave vector and the radial direction as $\omega\BV\sin\theta$, thus reducing the characteristic sloshing speed.

There are some further, expected differences between the toy model and the hydrodynamic simulations.
Over time, the sloshing process alters the entropy profile of the ICM in the cluster core, thus the motion of the sloshing fronts must change.
This aspect is not included in the toy model.
The effect is expected to be stronger for stronger perturbations, and later (i.e.~more inner) SCFs, due to a stronger resulting modification of the central entropy profile.
This is indeed the case.

The version of the toy model presented here considers only instantaneous perturbations, either simultaneously throughout the whole cluster, or a locally instantaneous perturbation moving through the cluster at a speed considerably larger than the characteristic sloshing speed.
We have also separated kick and offset perturbations in this version of the toy model.
A binary merger causes a more complex perturbation, extended in time and space, so a perfect match cannot be expected.

The current toy model only predicts speeds and locations of cold fronts, but not their strength in terms of density or temperature contrast across them, apart from predicting potentially `failed' outer cold fronts.

%*************
\subsection{Implications for determining merger ages}
There are two difficulties in deriving the age of a merger from an observation of a set of sloshing cold fronts in a given cluster.
Most clusters are observed first, and best, in their central regions, i.e.~we observe most easily the inner cold fronts of a sloshing front system.
If we misidentify them for outer cold fronts, we will generally overestimate their speed, and thus underestimate the age of the merger.
If only part of the sloshing front system is known, there is no easy way to know which cold fronts are observed.
Thus, merger ages from studies interpreting only sloshing in the cluster centre can only give a lower limit on the cluster's merger age (e.g.~\citealt{Roediger2011,Roediger2012a496}).

Even using a reasonable estimate of the particular sloshing front's speed, simply tracing back the current SCF radius to the cluster centre can strongly underestimate the age of the merger because the SCF emergence delay time of approximately $n\times 2/3\,$Gyr is not included.
We note that if the age of a particular front is of interest instead of the age of the merger, the simple trace-back method gives good estimates each sloshing front moves with approximately constant speed.
We note that the speed of a given cold front is approximately independent of the perturbation only for mild perturbations.
For, e.g., stronger mergers, the cold front speed increases \citep{Roediger2011,Bellomi2023}.

To estimate the age of the merger that caused the sloshing, we need a view of the cluster as a whole, and must identify the outermost sloshing front.
This front is affected least by the delay in emergence, and is a direct tracer of the merger age.
However, the search for the largest SCF must include looking for `failed' SCFs, and not simply use the outermost front with a discontinuity.
We show in a forthcoming paper that the outermost CF is indeed a good tracer of the merger's age.

The toy model implies a relationship between the orientation of the cold front system, and the merger direction.
However, even a single non-head-on merger introduces a rotational component into the ICM that rotates the sloshing direction, introducing a bias in direction.
This effect could depend on the impact parameter of the merger.

%*************
\subsection{Sloshing fronts as a wave phenomenon and their resilience against destruction by Kelvin-Helmholtz instability}
%Given that sloshing cold fronts are a wave phenomenon, they are no ordinary contact discontinuities but there is an underlying mechanism that re-establishes them continuously.
True SCFs are contact discontinuities with gas of different entropy on either side.
However, the identity of the gas on either side of a given front changes with time.
ICM that still is on the outside of a given front will be on its inside a while later when the front has moved further outwards.
SCFs are like waves moving through the ICM, they do not transport ICM from inner to outer radii over large distances.

Thus, SCFs differ in their nature from, e.g., the CF at the upstream edge of a subcluster falling into a host cluster.
In this scenario, the gas on the hotter side of the front is always host cluster ICM and the gas on the colder side always subcluster gas.
Gas parcels at this kind of cold front can be replaced by flows inside the subcluster or the flow of host cluster ICM around the subcluster atmosphere, but gas from each reservoir does not change to the other side of the front.
In contrast, at SCFs, material changes from the hotter side to the colder side as the sloshing front moves over it.

Thus, while we expect shear flows along sloshing fronts to cause Kelvin-Helmholtz instabilities (KHIs), we should not expect these KHIs to be able to fully erase a given SCF.
For example, KHIs of perturbation length 10\,kpc, arising on an interface of density contrast 2 and shear velocity 300\,km/s, have a growth time of 30\,Myr, but take about 5 growth times to form the classic KHI rolls (e.g.~\citealt{Roediger2013khi}), and would take even longer to erase the interface.
A naive interpretation could conclude that a $\sim$~Gyr old discontinuity should be erased by KHIs as their growth time is much shorter than the age of the front.
However, assuming a sloshing front speed of 0.05\,kpc/Myr (Figure~\ref{fig:CF_speeds_comparison}), in 150\,Myr, the sloshing front would travel 7.5\,kpc, i.e.~almost a perturbation length, which is typically larger than the KHI roll height.
Thus, KHI growth and sloshing front propagation, and re-formation, take place on similar timescales, which explains why KHIs generally cannot erase sloshing fronts.
Sloshing CFs are indeed known to survive KHIs from hydrodynamic simulations.
Rather than being erased or washed out, they are only distorted (e.g.~\citealt{Roediger2013virgovisc, ZuHone2013}).
Further ICM properties stabilising fronts against KHI, e.g.~viscosity (e.g.~\citealt{ZuHone2010,Roediger2013virgovisc}) or magnetic fields (e.g.~\citealt{Vikhlinin2002,Brzycki2019a,Chadayammuri2022a}), are not necessary to ensure front survival, but would help to slow down or even prevent the onset of KHIs.

%*********************************************************
\section{Conclusion and summary}
\label{sec_summary}
%*********************************************************
We presented a simple toy model for sloshing of the ICM in galaxy clusters that describes sloshing fronts as a coherent pattern arising from ICM parcels oscillating locally with their Brunt-Väisälä period.
This period can be approximated by a linear function of radius.
The proportionality constant, $1/u$, is the inverse of the characteristic speed of the resulting sloshing fronts, and is about 14\% of the ICM sound speed.
The simple model successfully predicts the staggered pattern of sloshing fronts on opposite sides of the cluster, the outwards motion of sloshing fronts with approximately constant speed, and the finite size of the sloshing front pattern.
Sloshing fronts should be numbered from the outside inwards.
A careful analysis of hydrodynamic simulations reveals that in the hydrodynamic treatment, sloshing fronts emerge near the cluster centre one after the other, with a delay of roughly 0.5\,Gyr between them.
This effect is not captured by the toy model.

We explained that the best option to derive the age of the merger that triggered sloshing is to trace back the outermost sloshing front.
However, such an analysis needs to take into account `failed' sloshing cold fronts, i.e.~those outer sloshing cold fronts that did not form a true discontinuity.
The existence of such `failed' sloshing cold fronts is predicted by the toy model.

\section*{Acknowledgements}
%The Acknowledgements section is not numbered. Here you can thank helpful
%colleagues, acknowledge funding agencies, telescopes and facilities used etc.
We thank the referee for their helpful comments.
IV acknowledges the support of University of Hull Astrophysical Data Science Cluster.
ER and IV acknowledge access to Viper, the University of Hull High Performance Computing Facility.
PN was supported under NASA contract NAS8-03060.

%%%%%%%%%%%%%%%%%%%%%%%%%%%%%%%%%%%%%%%%%%%%%%%%%%%
\section*{Data Availability}
The simulation data and code used for analysis from this study are available from ER upon reasonable request.
\bibliographystyle{mnras}
\bibliography{SloshingToyModel_arxiv.bib} % if your bibtex file is called example.bib
%%%%%%%%%%%%%%%%%%%%%%%%%%%%%%%%%%%%%%%%%%%%%%%%%%

%%%%%%%%%%%%%%%%% APPENDICES %%%%%%%%%%%%%%%%%%%%%

% Don't change these lines
\bsp	% typesetting comment
\label{lastpage}
\end{document}